\documentclass{article}

\usepackage{tikz, pgf, pgfplots}
\usetikzlibrary
	{
		arrows, automata, chains, datavisualization.formats.functions, fit, positioning, quantikz, quotes, shapes
	}
\usepackage{tikzpeople}
\usepackage{fontawesome5}

\usepackage{amsthm}
\usepackage{amssymb}
\usepackage[bb = boondox]{mathalfa}
\usepackage{dsfont}
\usepackage{mathtools}
\usepackage{multicol}
\usepackage{braket}
\usepackage{physics}
\numberwithin{equation}{section}

\usepackage{yquant, braket}

\usepackage[a4paper, left = 2.5cm, right = 2.5 cm, top = 2.5cm, bottom = 3.0 cm]{geometry}

\usepackage{xcolor}
\usepackage{varwidth}
\usepackage[breakable, skins, theorems]{tcolorbox}
\usepackage{graphicx}
\usepackage{hyperref}
\hypersetup
{
	colorlinks,
	linkcolor = {WordRed!75!black},
	citecolor = {WordBlueDarker25},
	urlcolor = {WordAquaDarker50}
}
\usepackage{nicematrix}
\NiceMatrixOptions
{
	code-for-first-row = \color{RedPurple},
	code-for-last-col = \color{WordAquaDarker25}
}
\usepackage{colortbl}
\usepackage{float}
\usepackage{cellspace}
\usepackage{makecell}
\usepackage[boxed, ruled, vlined]{algorithm2e}
\usepackage{appendix}
\usepackage{multirow}
\newcommand{\orcidicon}[1]{\href{https://orcid.org/#1}{\includegraphics[height=\fontcharht\font`\B]{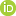}}}

\newtheorem{definition}{Definition}[section]

\definecolor{MyLightRed}{RGB}{244, 213, 245}
\definecolor{WordRed}{RGB}{255, 0, 102}
\definecolor{RedDarkLightest}{HTML}{ff0088}
\definecolor{RedDarkLight}{HTML}{ea005f}
\definecolor{RedPurple}{HTML}{aa007f}
\definecolor{Purple}{HTML}{911146}
\definecolor{WordLightGreen}{RGB}{140, 214, 192}
\definecolor{WordGreen}{RGB}{0, 176, 80}
\definecolor{GreenLightest}{HTML}{00ffa0}
\definecolor{GreenLighter1}{HTML}{00b383}
\definecolor{GreenLighter2}{HTML}{00aa7f}
\definecolor{GreenDark}{HTML}{225522}
\definecolor{GreenTeal}{HTML}{008080}
\definecolor{WordIceBlue}{RGB}{223, 227, 229}
\definecolor{MyVeryLightBlue}{RGB}{211, 245, 247}
\definecolor{WordBlueVeryLight}{RGB}{0, 176, 240}
\definecolor{WordBlueLight}{RGB}{0, 112, 192}
\definecolor{WordBlueDark}{RGB}{46, 116, 181}
\definecolor{WordBlueDarker}{RGB}{31, 78, 121}
\definecolor{WordBlueDarker25}{RGB}{54, 96, 146}
\definecolor{WordBlueDarker50}{RGB}{36, 64, 98}
\definecolor{WordBlueDarkest}{RGB}{0, 32, 96}
\definecolor{WordBlue}{RGB}{19, 65, 99}
\definecolor{MyBlue}{RGB}{0, 64, 128}
\definecolor{MyDarkBlue}{RGB}{0, 51, 102}
\definecolor{BlueVeryDark}{HTML}{222255}
\definecolor{WordAquaLighter80}{RGB}{218, 238, 243}
\definecolor{WordAquaLighter60}{RGB}{183, 222, 232}
\definecolor{WordAquaLighter40}{RGB}{146, 205, 220}
\definecolor{WordAquaDarker25}{RGB}{49, 134, 155}
\definecolor{WordAquaDarker50}{RGB}{33, 89, 103}
\definecolor{WordVeryLightTeal}{RGB}{223, 236, 235}
\definecolor{WordLightTeal}{RGB}{160, 199, 197}
\definecolor{WordDarkTealLighter80}{RGB}{207, 223, 234}
\definecolor{WordDarkTeal}{RGB}{72, 123, 119}
\definecolor{WordDarkerTeal}{RGB}{48, 82, 80}
\definecolor{WordTurquoiseLighter80}{RGB}{209, 238, 249}
\definecolor{Brown}{HTML}{666633}

\title
	{
		Quantum secret aggregation utilizing a network of agents
	}

\author
	{
		Michael Ampatzis$^1$\orcidicon{0000-0002-1570-6722}
		and
		Theodore Andronikos$^1$\orcidicon{0000-0002-3741-1271} \\
		$^1$Department of Informatics, Ionian University, \\
		7 Tsirigoti Square, 49100 Corfu, Greece; \\
		\{p16abat, andronikos\}@ionio.gr \\
	}

\begin{document}

\maketitle

\begin{abstract}
	In this work we consider the following problem: given a network of spies, all distributed in different locations in space, and assuming that each spy possesses a small, but incomplete by itself part of a big secret, is it possible to securely transmit all these partial secrets to the spymaster, so that they can be combined together in order to reveal the big secret? We refer to it as the Quantum Secret Aggregation problem, and we propose a protocol, in the form of a quantum game, with Alice taking over the role of the spymaster, that addresses this problem in complete generality. Our protocol relies on the use of maximally entangled GHZ tuples, which are symmetrically distributed among Alice and all her spies. It is the power of entanglement that makes possible the secure transmission of the small partial secrets from the agents to the spymaster. As an additional bonus, entanglement guarantees the security of the protocol, by making it statistically improbable for the notorious eavesdropper Eve to steal the big secret.
	\\
	\\
\textbf{Keywords:}: Quantum entanglement, GHZ states, quantum cryptography, quantum secret sharing, quantum secret aggregation.
\end{abstract}
\section{Introduction} \label{sec:Introduction}

Our rapidly growing dependence and continuous development of many prominent network-based technologies, such as the internet of things or cloud-based computing, have resulted, in an ever-growing need for more reliable and robust security protocols that can protect our current infrastructure from malicious individuals or parties. Even though our current security protocols, which base their security upon a set of computationally difficult mathematical problems like the factorization problem, have been proven reliable for the time being, they have also been proven vulnerable against more sophisticated attacks that incorporate the use of quantum algorithms and quantum computers. Despite the fact, that most of these quantum algorithms were theoretically developed a couple of decades ago, like the two famous algorithms developed by Peter Shor and Lov Grover \cite{Shor1994, Grover1996}, for many years, there was not any immediate threat of such attacks. That was simply because the technology of quantum computation was not mature enough to even produce a quantum computer capable of surpassing the $100$ qubit barrier, let alone of having the qubit capacity required to actually break these encryption protocols. 

However, after the monumental breakthrough of IBM's new quantum computers, which managed to surpass the $100$ qubit barrier \cite{IBMEagle} last year, and then immediately followed a year later by their most recent $433$ qubit quantum processor named Osprey \cite{IBMOsprey}, that managed to quadruple the previous processor's qubit capacity, the landscape has changed dramatically. It is now clear, that we are much closer to successfully developing a viable fully working quantum computer than originally anticipated. Thus, the need has arisen to immediate upgrade our security protocols, before they become a critical threat to our communication infrastructure. This inherent vulnerability of the current protocols has led to a plethora of initiatives from various countries and organizations, all aiming at establishing new and novel approaches for solving the ever-more critical problem of secure communication \cite{chamola2021information}. Among the various attempts to provide a viable solution for this problem, two new scientific fields emerged, namely the field of post-quantum or quantum-resistant cryptography and the field of quantum cryptography. Despite the confusing similarities in their names, these fields attempt to solve the problem by implementing radically different strategies. Specifically, the field of quantum-resistant cryptography is trying to maintain the philosophy of the previous era, by still relying on the use of mathematical problems, albeit of a more complex nature, such as supersingular elliptic curve isogenies and supersingular isogeny graphs, solving systems of multivariate equations, and lattice-based cryptography. On the other hand, the field of quantum cryptography is trying to establish security by relying on the fundamental principles of quantum mechanics, such the monogamy of entangled particles, the no-cloning theorem and nonlocality.

For the time being, due to the inherent restrictions of our current technology, the most prominent of the aforementioned fields is the field of post-quantum cryptography \cite{chen2016report,alagic2019status,alagic2020status,alagic2022status}, on account of the fact that the successful implementation of such protocols, does not require any changes of our current infrastructure. But in spite of all that, the field of quantum cryptography is still a crucial research topic, as it is wildly regarded as the long-term future of cryptography. This is due to the overwhelming advantages of the fundamental properties of quantum mechanics, which not only allow us to protect our information, but also efficiently transmit it using entangled states, as first proposed by Arthur Ekert \cite{Ekert1991}. In his E91 quantum key distribution protocol (QKD for short), Ekert proved that key distribution is possible with the use of EPR pairs. After this landmark discovery by Ekert, the field of quantum cryptography witnessed rapid growth in the development of entanglement-based QKD protocols \cite{Bennett1992,Gisin2004,inoue2002differential, guan2015experimental,waks2006security,Ampatzis2021}, thus proving the technique's importance, while simultaneously prompting the research community to expand the field by experimenting with other cryptographic primitives like quantum secret sharing.

The cryptographic primitive of secret sharing or secret splitting in its more elementary form can be described as a game between two spatially separated groups. The first group consisting of one player, who wants to share a secret message with the other group. This second group consists of the rest of the players, who will receive the secret message split into multiple pieces. By itself each piece does not contain any valuable information, but, if all the players of the second group were to combine their pieces, the secret message would be revealed. Understandably, one can make the mistake and regard this cryptographic primitive, as nothing more than a scaled-up key distribution protocol, in order to accommodate more than two people. However, the step of dividing the secret message into multiple pieces actually offers a crucial advantage, by providing security against malicious individuals that have managed to infiltrate the second group with the goal of covertly acquiring the secret message, by forcing every player honest or dishonest to participate in the process that unlocks the secret message (see the recent \cite{Ampatzis2022} for more details).

In the real world, secret sharing schemes are vital for providing security to new and emerging technologies, such as the fields of cloud computing, cloud storage \cite{attasena2017secret,ermakova2013secret} or blockchain technologies \cite{cha2021blockchain}. These technologies require multiple parties to communicate with each other, accommodating the possibility that one or more of them might be malicious users, who want to take advantage of the system. Therefore, the research on quantum secret sharing has come a long way from the simple proof of concept by Hillery et al. \cite{hillery1999quantum}, and Cleve et al. \cite{cleve1999share}, who pioneered this field. All this progress has led to numerous research proposals and schemes that are continuously expanding the field to this day \cite{karlsson1999quantum,smith2000quantum,gottesman2000theory,fortescue2012reducing,qin2020hierarchical,senthoor2022theory}. At the same time, multiple experimental demonstrations involving real-world scenarios have been attempted by the researchers in \cite{fu2015long,wu2020passive,grice2019quantum,gu2021secure}, and even schemes for non-binary quantum secret sharing protocols that rely on the use of qudits instead of qubits \cite{keet2010quantum,helwig2012absolute,liu2018quantum,mansour2020quantum} have been proposed.

This work tackles a problem that could be considered the inverse of quantum secret sharing. In our setting, there is a network of agents, all distributed in different locations, and all in possession of a small secret. All these small secrets must be combined together, if one is to reveal the big secret. So, the agents want to transmit their secrets to the spymaster, who is located elsewhere. Our agents operate on a need to know basis, that is they avoid any communication among them, and only report directly to the spymaster. Their task is complicated by the need to securely fulfill their mission, as adversaries might try to intercept any message and discover the big secret. Thus, going quantum seems the way to go. We refer to this problem as Quantum Secret Aggregation, and we give a protocol that solves this task in the form of a game. The use of games does not diminish the seriousness or importance of the problem, but, at least we hope so, makes its presentation more entertaining and memorable. Certainly this is not the first time games, coin tossing, etc. have been used in quantum cryptography (see \cite{Bennett1984} and recently \cite{Ampatzis2021,Ampatzis2022}). Quantum games have captured the interest of many researchers since their inception in 1999 \cite{Meyer1999}, \cite{Eisert1999}. In many situations, quantum strategies seem to outperform classical ones \cite{Andronikos2018,Andronikos2021,Andronikos2022a}. This holds not only for iconic classical games like the Prisoners' Dilemma \cite{Eisert1999}, \cite{Giannakis2019}, but also for abstract quantum games \cite{Giannakis2015a}. As a matter of fact, there is no inherent restriction on the type of a classical system that can be transformed to a quantum analogue, as even political institutions may be amenable to this process \cite{Andronikos2022}. In closing this short reference, it is perhaps noteworthy that many games have been studied via unconventional means outside the quantum realm. The realization that nature computes has also been applied to bio-molecular processes (see for instance \cite{Theocharopoulou2019,Kastampolidou2020a,Kostadimas2021}). It should therefore come as no surprise that, in order to improve classical strategies in many famous games, including the Prisoners' Dilemma, tools and techniques from the biological domain have been utilized \cite{Kastampolidou2020,Kastampolidou2021,Kastampolidou2021a,Papalitsas2021}.

\textbf{Contribution}. This paper poses and solves a problem in the general context of quantum cryptographic protocols. We refer to it as the Quantum Secret Aggregation problem because it involves aggregating many small secrets in order to reveal a big secret. The underlying setting visualizes a completely distributed network of agents, each in possession of a small secret, aiming to send their secret to the spatially separated Alice, which is our famous spymaster. The operation must be completed in the most secure way possible, as there are eavesdroppers eager to intercept their communications and steal the big secret. To address this problem we present the Quantum Secret Aggregation Protocol as a game. The solution outlined is completely general, as the number of players can be scaled arbitrarily as needed and all $n$ players are assumed to reside in different positions in space. Obviously, the solution still holds, even if a subset of the players are located in the same place. Security is enforced because of the integral role of entanglement in the protocol. The use of maximally entangled GHZ tuples, symmetrically distributed among Alice and all her spies not only makes possible the secure transmission of the small partial secrets from the agents to Alice, but also guarantees the security of the protocol, by making it statistically improbable for the notorious eavesdropper Eve to obtain the big secret.

\subsection{Organization} \label{subsec:Organization}

The structure of this paper is the following. Section \ref{sec:Introduction} provides an introduction to the subject along with some relevant references. Section \ref{sec:Brief Reminder About GHZ States} is a brief exposition on GHZ states and the phenomenon of entanglement. Section \ref{sec:The Quantum Secret Aggregation Problem} rigorously defines the problem at hand, while Section \ref{sec:The Quantum Secret Aggregation Protocol} explains in detail the Quantum Secret Aggregation protocol. Section \ref{sec:QSA Protocol Example} presents a small example of the protocol executed using Qiskit. Section \ref{sec:Security Analysis of the QSA Protocol} is devoted to the security analysis on a number of possible attacks from, and, finally, Section \ref{sec:Discussion and Conclusions} contains a summary and a brief discussion on some of the finer points of this protocol.

\section{A brief reminder about GHZ states} \label{sec:Brief Reminder About GHZ States}

Nowadays most quantum protocols designed to securely transmit keys, secrets, and information in general, rely on the power of entanglement. Entanglement is a hallmark property of the quantum world. As this phenomenon is absent from the everyday world, it is considered counterintuitive by some. However, from the point of view of quantum cryptography and quantum computation, this strange behavior is seen as a precious resource, which is the key to achieve quantum teleportation and unconditionally secure information transmission.

Thus, it comes as no surprise that this work too utilizes quantum entanglement in a critical manner, in order to implement the proposed protocol of quantum secret aggregation. Specifically, our protocol relies on maximally entangled $n$-tuples of qubits, i.e., qubits that are in what in the literature is reffered to as the $\ket{ GHZ_{n} }$ state. Present-day quantum computers can produce arbitrary GHZ states using various quantum circuits. A methodology for constructing efficient such circuits is given in \cite{Cruz2019}. The resulting quantum circuits are efficient in the sense that they require $\lg n$ steps to generate the $\ket{ GHZ_{n} }$ state. One such circuit that generates the $\ket{ GHZ_{5} }$ state using the IBM Quantum Composer \cite{IBMQuantumComposer2022} is shown in Figure \ref{fig:GHZ_5_QC}. The dotted lines are a helpful visualization that allows us to distinguish ``time slices'' within which the CNOT gates are applied in parallel. Figure \ref{fig:GHZ_5_SV}, which is also from the IBM Quantum Composer, indicates the state vector description of the $\ket{ GHZ_{5} }$ state.

Let us assume that we are given a composite quantum system made up of $n$ individual subsystems, where each subsystem contains just a single qubit. As explained above, it is possible to entangle all these $n$ of the composite system qubits in the $\ket{ GHZ_n }$ state. In such a case, the mathematical description of the state of the composite system is the following:

\begin{align} \label{eq:Extended General GHZ_n State}
	\ket{ GHZ_{n} }
	=
	\frac{ 1 }{ \sqrt{2} }
	\left( \ket{0}_{ n - 1 } \ket{0}_{ n - 2 } \dots \ket{0}_{ 0 } + \ket{1}_{ n - 1 } \ket{1}_{ n - 2 } \dots \ket{1}_{ 0 } \right)
	\ ,
\end{align}

where the subscript $i, \ 0 \leq i \leq n - 1,$ is used to indicate the qubit belonging to subsystem $i$.

\begin{tcolorbox}
	[
		grow to left by = 0.0 cm,
		grow to right by = 0.0 cm,
		colback = WordTurquoiseLighter80!07,	
		enhanced jigsaw,						
		sharp corners,
		toprule = 1.0 pt,
		bottomrule = 1.0 pt,
		leftrule = 0.1 pt,
		rightrule = 0.1 pt,
		sharp corners,
		center title,
		fonttitle = \bfseries
	]
	\begin{figure}[H]
		\centering
		\includegraphics[scale = 0.50, trim = {0 0cm 0cm 0}, clip]{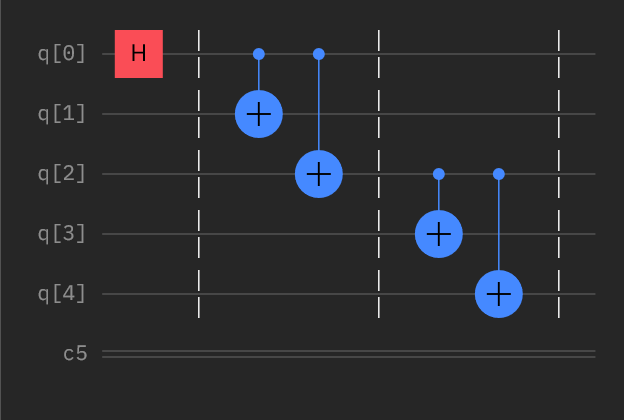}
		\caption{The above (efficient) quantum circuit in Qiskit can entangle 5 qubits in the $\ket{ GHZ_5 } = \frac{ \ket{0} \ket{0} \ket{0} \ket{0} \ket{0} + \ket{1} \ket{1} \ket{1} \ket{1} \ket{1} } {\sqrt{2}}$ state. Following the same pattern, we can be construct efficient quantum circuits that entangle $n$ qubits in the  $\ket{ GHZ_n }$ state.}
		\label{fig:GHZ_5_QC}
	\end{figure}
\end{tcolorbox}

\begin{tcolorbox}
	[
		grow to left by = 0.0 cm,
		grow to right by = 0.0 cm,
		colback = white,			
		enhanced jigsaw,			
		sharp corners,
		toprule = 1.0 pt,
		bottomrule = 1.0 pt,
		leftrule = 0.1 pt,
		rightrule = 0.1 pt,
		sharp corners,
		center title,
		fonttitle = \bfseries
	]
	\begin{figure}[H]
		\centering
		\includegraphics[scale = 0.375, trim = {0cm 4.25cm 10cm 0cm}, clip]{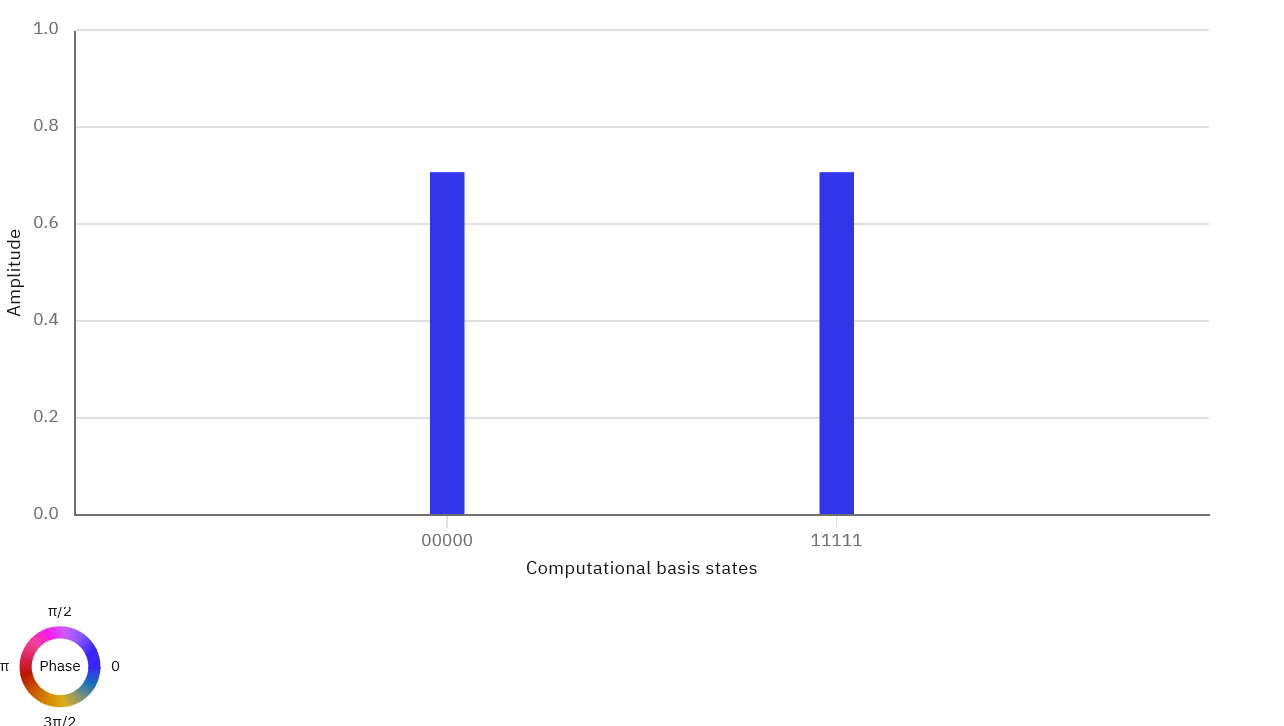}
		\caption{This figure depicts the state vector description of 5 qubits that are entangled in the $\ket{ GHZ_5 }$ state.}
		\label{fig:GHZ_5_SV}
	\end{figure}
\end{tcolorbox}

It is expedient and necessary to generalize the above setting to the case where each individual subsystem is made of a quantum register and not just a single qubit. In this more general situation, each of the $n$ subsystems is a quantum register $r_i$, where $0 \leq i \leq n - 1,$ that has $m$ qubits and the \emph{corresponding} qubits of all the $n$ registers are entangled in the $\ket{ GHZ_{n} }$ state. This means that all the qubits in position $j, \ 0 \leq j \leq m - 1,$ of the registers $r_0, r_1 \dots, r_{ n - 1 }$ are entangled in the $\ket{ GHZ_{n} }$ state. Figure \ref{fig:Composite System of $n$ Registers} provides a visual depiction of this situation, where the corresponding qubits comprising the $\ket{ GHZ_{n} }$ $n$-tuple are drawn with the same color. As expected, the state of the composite system is designated by $\ket{ GHZ_{n} }^{\otimes m}$ and its mathematical description is

\begin{align} \label{eq:m Extended General GHZ_n States}
	\ket{ GHZ_{n} }^{\otimes m}
	&=
	\frac{1}{ \sqrt{2^m} }
	\sum_{\mathbf{x} \in \{ 0, 1 \}^m}
	\ket{\mathbf{x}}_{ n - 1 } \dots \ket{\mathbf{x}}_{ 0 }
	\ ,
\end{align}

where ${\mathbf{x} \in \{ 0, 1 \}^m}$ ranges through all the $2^{m}$ basis kets.

\begin{align} \label{eq: m+1 GHZ_n states}
	\hspace{- 1.00 cm}
	\ket{GHZ_{n}}^{\otimes m + 1}
	&=
	\ket{GHZ_{n}}^{\otimes m}
	\otimes
	\ket{GHZ_{n}}
	\nonumber \\
	&=
	\frac{1}{ \sqrt{2^m} }
	\sum_{\mathbf{x} \in \{ 0, 1 \}^m}
	\ket{\mathbf{x}}_{ n - 1 } \dots \ket{\mathbf{x}}_{ 0 }
	\otimes
	\frac{ 1 }{ \sqrt{2} }
	\left( \ket{0}_{ n - 1 } \ket{0}_{ n - 2 } \dots \ket{0}_{ 0 } + \ket{1}_{ n - 1 } \ket{1}_{ n - 2 } \dots \ket{1}_{ 0 } \right)
	\nonumber \\
	&=
	\frac{1}{ \sqrt{2^{m + 1}} }
	\sum_{\mathbf{x} \in \{ 0, 1 \}^m}
	\ket{ \mathbf{x} 0 }_{ n - 1 } \dots \ket{ \mathbf{x} 0 }_{ 0 }
	+
	\ket{ \mathbf{x} 1 }_{ n - 1 } \dots \ket{ \mathbf{x} 1 }_{ 0 }
	\nonumber \\
	&=
	\frac{1}{ \sqrt{2^{m + 1}} }
	\sum_{\mathbf{x} \in \{ 0, 1 \}^{m + 1}}
	\ket{ \mathbf{x} }_{ n - 1 } \dots \ket{ \mathbf{x} }_{ 0 }
	\ .
\end{align}

\begin{tcolorbox}
	[
		grow to left by = 2.0 cm,
		grow to right by = 0.0 cm,
		colback = yellow!03!white,	
		enhanced jigsaw,					
		sharp corners,
		toprule = 1.0 pt,
		bottomrule = 1.0 pt,
		leftrule = 0.1 pt,
		rightrule = 0.1 pt,
		sharp corners,
		center title,
		fonttitle = \bfseries
	]
	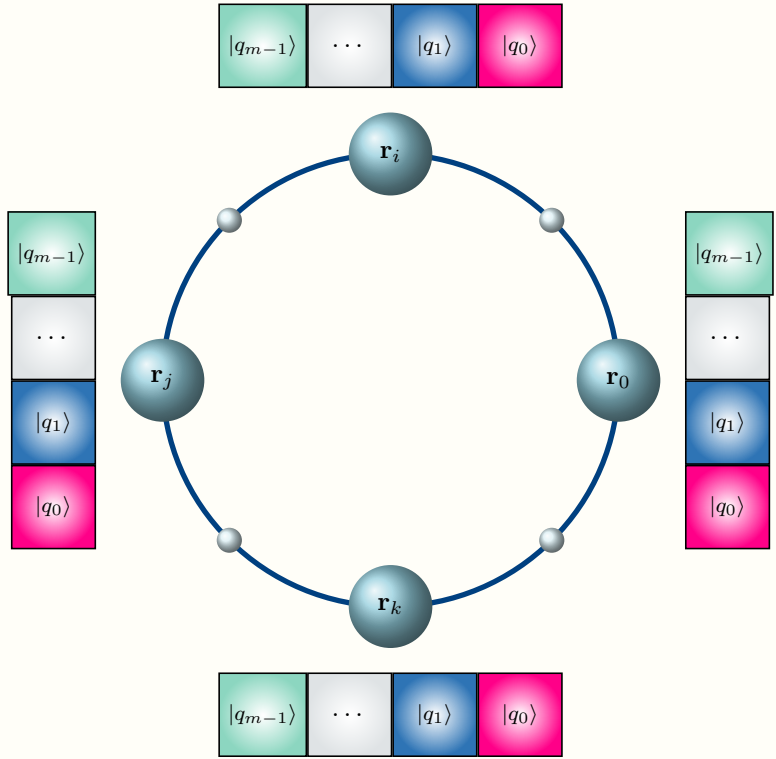
\begin{figure}[H]
		\centering
		\begin{tikzpicture} [ scale = 0.75 ]
			\def \n {8}
			\def \Angle {360 / \n}
			\def \InnerRadius {4}
			\def \OuterRadius {5}
			\node [ shade, top color = RedPurple, bottom color = black, rectangle, align = center, text width = 10.00 cm ] ( Label ) at ( 0.0, 9.00 )
			{ \color{white} \textbf{A composite system consisting of $n$ quantum registers $r_{0}, \dots, r_{n - 1}$. Each register has $m$ qubits and the corresponding qubits are entangled in the $\ket{ GHZ_n }$ state.} };
			\draw [ line width = 2.00 pt, MyBlue ] ( 0, 0 ) circle [ radius = \InnerRadius ];
			\node [ shade, shading = ball, ball color = WordAquaLighter40, circle, minimum size = 11 mm ] ( r0 ) at ( { \InnerRadius * cos(0 * \Angle) }, { \InnerRadius * sin(0 * \Angle) } ) { $\mathbf{r}_{0}$ };
			\matrix
			[
			anchor = west,
			column sep = 0.00 mm,
			row sep = 0.0 mm,
			]
			at ( { \OuterRadius * cos(0 * \Angle) }, { \OuterRadius * sin(0 * \Angle) } )
			{
				\node
				[ draw = black, shade, outer color = WordLightGreen, inner color = white, minimum size = 11 mm, semithick ] { \footnotesize $\ket{ q_{m - 1} }$ };
				\\
				\node
				[ draw = black, shade, outer color = WordIceBlue, inner color = white, minimum size = 11 mm, semithick ] { \dots };
				\\
				\node
				[ draw = black, shade, outer color = WordBlueDark, inner color = white, minimum size = 11 mm, semithick ] { \footnotesize $\ket{ q_{1} }$ };
				\\
				\node
				[ draw = black, shade, outer color = RedDarkLightest, inner color = white, minimum size = 11 mm, semithick ] { \footnotesize $\ket{ q_{0} }$ };
				\\
			};
			\node [ shade, shading = ball, ball color = WordAquaLighter80, circle ] () at ( { \InnerRadius * cos(1 * \Angle) }, { \InnerRadius * sin(1 * \Angle) } ) {};
			\node [ shade, shading = ball, ball color = WordAquaLighter40, circle, minimum size = 11 mm ] ( ri ) at ( { \InnerRadius * cos(2 * \Angle) }, { \InnerRadius * sin(2 * \Angle) } ) { $\mathbf{r}_{i}$ };
			\matrix
			[
			anchor = south,
			column sep = 0.000 mm, 
			row sep = 0.0 mm,
			]
			at ( { \OuterRadius * cos(2 * \Angle) }, { \OuterRadius * sin(2 * \Angle) } )
			{
				\node
				[ draw = black, shade, outer color = WordLightGreen, inner color = white, minimum size = 11 mm, semithick ] { \footnotesize $\ket{ q_{m - 1} }$ };
				&
				\node
				[ draw = black, shade, outer color = WordIceBlue, inner color = white, minimum size = 11 mm, semithick ] { \dots };
				&
				\node
				[ draw = black, shade, outer color = WordBlueDark, inner color = white, minimum size = 11 mm, semithick ] { \footnotesize $\ket{ q_{1} }$ };
				&
				\node
				[ draw = black, shade, outer color = RedDarkLightest, inner color = white, minimum size = 11 mm, semithick ] { \footnotesize $\ket{ q_{0} }$ };
				\\
			};
			\node [ shade, shading = ball, ball color = WordAquaLighter80, circle ] () at ( { \InnerRadius * cos(3 * \Angle) }, { \InnerRadius * sin(3 * \Angle) } ) {};
			\node [ shade, shading = ball, ball color = WordAquaLighter40, circle, minimum size = 11 mm ] ( rj ) at ( { \InnerRadius * cos(4 * \Angle) }, { \InnerRadius * sin(4 * \Angle) } ) { $\mathbf{r}_{j}$ };
			\matrix
			[
			anchor = east,
			column sep = 0.00 mm,
			row sep = 0.0 mm,
			]
			at ( { \OuterRadius * cos(4 * \Angle) }, { \OuterRadius * sin(4 * \Angle) } )
			{
				\node
				[ draw = black, shade, outer color = WordLightGreen, inner color = white, minimum size = 11 mm, semithick ] { \footnotesize $\ket{ q_{m - 1} }$ };
				\\
				\node
				[ draw = black, shade, outer color = WordIceBlue, inner color = white, minimum size = 11 mm, semithick ] { \dots };
				\\
				\node
				[ draw = black, shade, outer color = WordBlueDark, inner color = white, minimum size = 11 mm, semithick ] { \footnotesize $\ket{ q_{1} }$ };
				\\
				\node
				[ draw = black, shade, outer color = RedDarkLightest, inner color = white, minimum size = 11 mm, semithick ] { \footnotesize $\ket{ q_{0} }$ };
				\\
			};
			\node [ shade, shading = ball, ball color = WordAquaLighter80, circle ] () at ( { \InnerRadius * cos(5 * \Angle) }, { \InnerRadius * sin(5 * \Angle) } ) {};
			\node [ shade, shading = ball, ball color = WordAquaLighter40, circle, minimum size = 11 mm ] ( rk ) at ( { \InnerRadius * cos(6 * \Angle) }, { \InnerRadius * sin(6 * \Angle) } ) { $\mathbf{r}_{k}$ };
			\matrix
			[
			anchor = north,
			column sep = 0.000 mm, 
			row sep = 0.0 mm,
			nodes = { draw = black, shade, outer color = WordLightGreen, inner color = white, minimum size = 11 mm, semithick }
			]
			at ( { \OuterRadius * cos(6 * \Angle) }, { \OuterRadius * sin(6 * \Angle) } )
			{
				\node
				[ draw = black, shade, outer color = WordLightGreen, inner color = white, minimum size = 11 mm, semithick ] { \footnotesize $\ket{ q_{m - 1} }$ };
				&
				\node
				[ draw = black, shade, outer color = WordIceBlue, inner color = white, minimum size = 11 mm, semithick ] { \dots };
				&
				\node
				[ draw = black, shade, outer color = WordBlueDark, inner color = white, minimum size = 11 mm, semithick ] { \footnotesize $\ket{ q_{1} }$ };
				&
				\node
				[ draw = black, shade, outer color = RedDarkLightest, inner color = white, minimum size = 11 mm, semithick ] { \footnotesize $\ket{ q_{0} }$ };
				\\
			};
			\node [ shade, shading = ball, ball color = WordAquaLighter80, circle ] () at ( { \InnerRadius * cos(7 * \Angle) }, { \InnerRadius * sin(7 * \Angle) } ) {};
		\end{tikzpicture}
		\caption{This figure visualizes the situation where each of the $n$ subsystems is a quantum register $r_i, \ 0 \leq i \leq n - 1,$ that has $m$ qubits, and the corresponding qubits in all the registers, drawn above with the same color, are entangled in the $\ket{ GHZ_{n} }$ state.} \label{fig:Composite System of $n$ Registers}
	\end{figure}
\end{tcolorbox}

Equation (\ref{eq:m Extended General GHZ_n States}) can be proved by an easy induction on $m$. For $m = 1$, equation (\ref{eq:m Extended General GHZ_n States}) reduces to (\ref{eq:Extended General GHZ_n State}), and trivially holds. Let us assume that, according to the induction hypothesis, (\ref{eq:m Extended General GHZ_n States}) holds for $m$. We shall prove that (\ref{eq:m Extended General GHZ_n States}) also holds for $m + 1$. Indeed, by invoking (\ref{eq:Extended General GHZ_n State}) and (\ref{eq:m Extended General GHZ_n States}), the computation shown below completes the proof by induction.

\section{The problem of Quantum Secret Aggregation} \label{sec:The Quantum Secret Aggregation Problem}

In the current section we rigorously define the problem of \emph{Quantum Secret Aggregation}, simply referred to as QSA from now on. To the best of our knowledge, this is the first time that this problem is posed and solved in the relevant literature. Informally, QSA can be considered as the inverse of Quantum Secret Sharing (QSS for short). The latter focuses on how a single entity (usually called Alice) can securely transmit a secret to a group of two or more agents. Typically in QSS Alice is in a different location from her agents; however the agents are assumed to be in the same location, which implies that they can readily exchange information. In contrast, in QSA we assume that Alice and her agents are all in different locations, and this time it is the agents that want to securely transmit a part of the secret to Alice. Each agent has only a small part of the secret, and no two agents possess secrets with common fragments. Alice requires all the parts in order to decipher the secret.

\begin{definition} [Quantum Secret Aggregation] \label{def:Quantum Secret Aggregation}
	Let us assume that the following hold.

	\renewcommand\labelenumi{(\textbf{A}$_\theenumi$)}
	\begin{enumerate}
		\item	There are $n - 1$ \emph{spatially separated} agents Agent$_0$, \dots, Agent$_{n - 2}$. Each agent possesses of a \emph{partial secret key} $\mathbf{p}_i, \ 0 \leq i \leq n - 2$.
		\item	Every partial secret keys is unique and is known only to the corresponding agent. Furthermore, there is no information redundancy among the partial secret keys, i.e., no one can be inferred from the rest.
		\item	Every agent wants to securely send her secret key to the spymaster Alice, who is also in an entirely different location.
		\item	Alice wants to discover the \emph{complete secret key}, denoted by $\mathbf{s}$. This can only be done by combining all the partial secret keys $\mathbf{p}_0, \dots, \mathbf{p}_{n - 2}$.
		\item	The length of the complete secret key, denoted by $m$, is the sum of the lengths of all the partial secret keys: $m = | \mathbf{p}_0 | + \dots + | \mathbf{p}_{n - 2} |$.
		\item	The whole operation must be executed with utmost secrecy, due to the presence of the eavesdropper Eve.
	\end{enumerate}

	The \emph{Quantum Secret Aggregation} problem asks how to establish a protocol that will guarantee that Alice and her agents achieve their goal.
\end{definition}

In view of the fact that Agent$_i$ possesses the partial key $\mathbf{p}_i, \ 0 \leq i \leq n - 2$, we can make the following observations.

\begin{itemize}
	\item	Implicit in the definition of the problem is the assumption that Alice has assigned a specific ordering to her ring of agents and all her agents are aware of this ordering. This simply means that not only Alice, but also all agents know who is Agent$_0$, \dots, Agent$_{n - 2}$.
	\item	Definition \ref{def:Quantum Secret Aggregation} is general enough to allow for partial secret keys of different length, which is more realistic.
	\item	Although neither Alice nor her agents know the partial secret keys (except their own),  they all know their lengths $| \mathbf{p}_0 |, \dots, | \mathbf{p}_{n - 2} |$. This does not compromise the secrecy factor because knowing the length of a secret key does not reveal its contents.
\end{itemize}

From an algorithmic perspective it is convenient to have a standard length for all partial secret keys. This prompts the following definition.

\begin{definition} [Extended Partial Secret Key] \label{def:Extended Partial Secret Key}
	Each Agent$_i, \ 0 \leq i \leq n - 2,$ constructs from her partial secret key $\mathbf{p}_i$ her \emph{extended} partial secret key $\mathbf{s}_i$, which is defined as
	\begin{align} \label{eq:Extended Partial Key}
		\mathbf{s}_i
		=
		\underbrace{ 0 \ \cdots \ 0 }_{k \rm\ times}
		\ \mathbf{p}_i \
		\underbrace{ 0 \ \cdots \ 0 }_{l \rm\ times}
		\ ,
	\end{align}
	where $k = | \mathbf{p}_{n - 2} | + \dots + | \mathbf{p}_{i + 1} |$ and $l = | \mathbf{p}_{i - 1} | + \dots + | \mathbf{p}_0 |$.
\end{definition}

This simple construction enforces uniformity among the agents, since they all end up having extended keys of length $m$, even though their partial keys will in general be of different length, and greatly simplifies the construction of the quantum circuit. Additionally, it enables us to derive the next simple and elegant formula connecting the complete secret key $\mathbf{s}$ with the 
extended partial secret keys $\mathbf{s}_{0}, \dots, \mathbf{s}_{n - 2}$:
\begin{align} \label{eq:Complete Secret as Mod $2$ Sum of Partial Keys}
	\mathbf{s}
	=
	\mathbf{s}_{0} \oplus \dots \oplus \mathbf{s}_{n - 2}
	\ .
\end{align}

\section{The Quantum Secret Aggregation protocol} \label{sec:The Quantum Secret Aggregation Protocol}

We know present the proposed QSA protocol as a game, aptly named the QSA game. In this game, there are $n, n \geq 3,$ players, which can be conceptually divided in two group. Alice alone makes the first group, which is the recipient of the secret information from distant sources. These sources are the $n - 1$ agents in the spy ring that constitute the second group. The proposed protocol is general enough to accommodate an arbitrary number of agents. To thoroughly describe the QSA game, we carefully distinguish the phases in its progression.

\begin{tcolorbox}
	[
		grow to left by = 0.00 cm,
		grow to right by = 0.00 cm,
		colback = yellow!03!white,			
		enhanced jigsaw,					
		sharp corners,
		toprule = 1.0 pt,
		bottomrule = 1.0 pt,
		leftrule = 0.1 pt,
		rightrule = 0.1 pt,
		sharp corners,
		center title,
		fonttitle = \bfseries
	]
	\begin{figure}[H]
		\centering
		\begin{tikzpicture} [scale = 1.00]
			\def \n {8}
			\def \Angle {360 / \n}
			\def \Radius {4}
			\def \Begin {0.4}
			\def \End {3.6}
			\node
			[
			shade, top color = GreenLighter2, bottom color = black, rectangle, text width = 9.00 cm, align = center
			] ( Label ) at ( 0.0, 7.00 )
			{ \color{white} \textbf{THE QUANTUM CHANNEL}\\
				Alice sends through the quantum channel $m$ qubits in the $\ket{ GHZ_{n} }$ state to each of the $n - 1$ spatially distributed agents in her spy network.};
			\draw [ line width = 2.00 pt, MyBlue ] ( 0, 0 ) circle [ radius = \Radius cm ];
			\node
			[
			maninblack,
			scale = 1.50,
			anchor = center,
			label = { [ label distance = 0.00 cm ] east: \textbf{Agent$_0$} }
			]
			( ) at ( { \Radius * cos(0 * \Angle) }, { \Radius * sin(0 * \Angle) } ) { };
			\node
			[
			maninblack, shirt = WordBlueDark, hair = gray,
			scale = 1.50,
			anchor = center,
			label = { [ rotate = \Angle, label distance = 0.00 cm ] east: \textbf{Agent$_1$} }
			]
			( ) at ( { \Radius * cos(1 * \Angle) }, { \Radius * sin(1 * \Angle) } ) { };
			\node
			[
			maninblack, shirt = GreenTeal, hair = red!20!black,
			scale = 1.50,
			anchor = center,
			label = { [ label distance = 0.00 cm ] north: \textbf{Agent$_2$} }
			]
			( ) at ( { \Radius * cos(2 * \Angle) }, { \Radius * sin(2 * \Angle) } ) { };
			\node [ shade, shading = ball, ball color = WordAquaLighter80, circle ] () at ( { \Radius * cos(3 * \Angle) }, { \Radius * sin(3 * \Angle) } ) {};
			\node
			[
			maninblack, shirt = brown!50!gray, skin = brown!70!black,
			scale = 1.50,
			anchor = center,
			label = { [ label distance = 0.00 cm ] west: \textbf{Agent$_i$} }
			]
			( ) at ( { \Radius * cos(4 * \Angle) }, { \Radius * sin(4 * \Angle) } ) { };
			\node [ shade, shading = ball, ball color = WordAquaLighter80, circle ] () at ( { \Radius * cos(5 * \Angle) }, { \Radius * sin(5 * \Angle) } ) {};
			\node
			[
			maninblack, shirt = red!50!blue, hair = brown!30!black, skin = brown,
			scale = 1.50,
			anchor = center,
			label = { [ label distance = 0.00 cm ] south: \textbf{Agent$_{n - 3}$} }
			]
			( ) at ( { \Radius * cos(6 * \Angle) }, { \Radius * sin(6 * \Angle) } ) { };
			\node
			[
			maninblack, shirt = green!20!blue, hair = gray!70!black, skin = brown,
			scale = 1.50,
			anchor = center,
			label = { [ rotate = - \Angle, label distance = 0.00 cm ] right: \textbf{Agent$_{n - 2}$} }
			]
			( ) at ( { \Radius * cos(7 * \Angle) }, { \Radius * sin(7 * \Angle) } ) { };
			\node
			[
			alice,
			scale = 1.50,
			anchor = center,
			label = { [ label distance = 0.00 cm ] south: \textbf{Alice} }
			]
			(Alice) at (0.0, 0.0) { };
			\begin{scope}[on background layer]
				\foreach \angle in { 0, 45, 90, 270, 315 }
				\draw [ WordAquaLighter40, ->, >=stealth, line width = 5.0 mm ] 
				( { \Begin * cos(\angle) }, { \Begin * sin(\angle) } ) --
				( { \End * cos(\angle) }, { \End * sin(\angle) } )
				node [ midway, text = black, rotate = \angle ] {$\ket{ GHZ_{n} }^{\otimes m}$};
				\foreach \angle in { 135, 225 }
				\node [ shade, shading = ball, ball color = WordAquaLighter80, circle ] () at ( { \Radius * 0.5 * cos(\angle) }, { \Radius * 0.5 * sin(\angle) } ) {};
				\draw [ WordAquaLighter40, ->, >=stealth, line width = 5.0 mm ]
				( { \Begin * cos(180) }, { \Begin * sin(180) } ) --
				( { \End * cos(180) }, { \End * sin(180) } )
				node [ midway, text = black ] {$\ket{ GHZ_{n} }^{\otimes m}$} ;
			\end{scope}
		\end{tikzpicture}
		\caption{The above figure depicts the situation where Alice herself initiates the protocol by creating and sending through the quantum channel to each of the $n - 1$ spatially distributed agents in her spy network $m$ qubits entangled in the $\ket{ GHZ_{n} }$ state.} \label{fig:Alice Transmits GHZ_n Tuples}
	\end{figure}
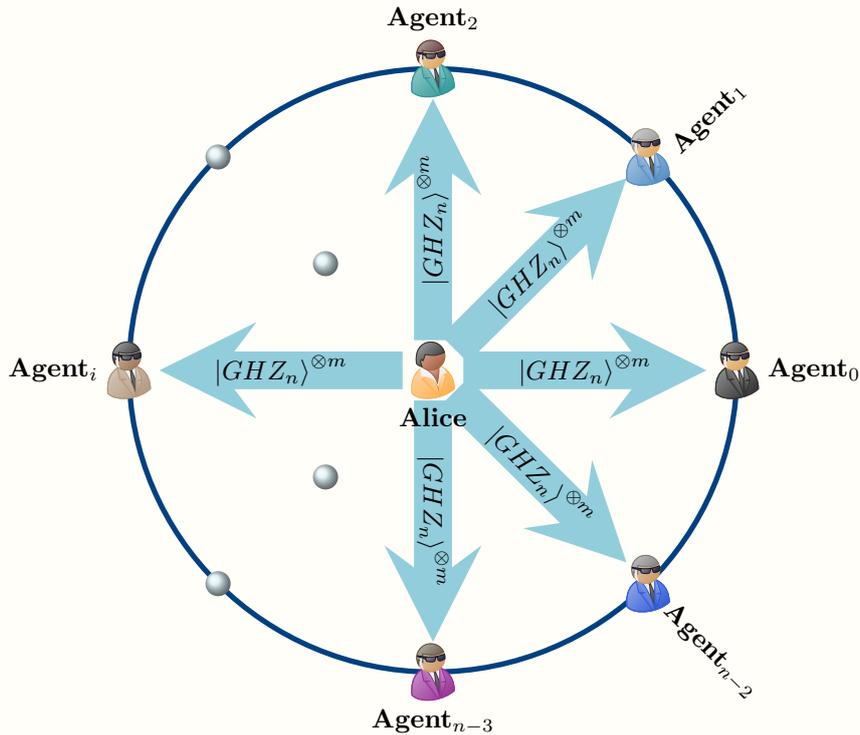
\end{tcolorbox}

\subsection{Initialization phase through the quantum channel} \label{sec:Initialization Phase Through the Quantum Channel}

This game utilizes entanglement. As a matter of fact, its successful completion relies on the use of entanglement. So, it is necessary, before the main part of the protocol commences, to create the required number, which is denoted by $m$, of $n$-tuples of qubits entangled in the $\ket{ GHZ_{n} }$ state. Such entangled tuples can be produced by a contemporary quantum computer, for instance, using a quantum circuit like the one shown in Figure \ref{fig:GHZ_5_QC}. These $\ket{ GHZ_{n}}$ tuples can be produced by Alice or by another trusted source, which can even be a satellite \cite{aspelmeyer2003long}. Figure \ref{fig:Alice Transmits GHZ_n Tuples} depicts the former situation. We note however that our protocol does not depend on which source actually creates the entangled tuples. The crucial requirement is that they are produced and sent through the quantum channel, so that they may populate the Input Registers of Alice and all her agents.

\subsection{Input phase in the local quantum circuits} \label{subsec:Input Phase in the Local Quantum Circuits}

The purpose of the QSA game from Alice's point of view is to aggregate all the partial secret keys $\mathbf{p}_0, \dots, \mathbf{p}_{n - 2}$ from her $n - 1$ agents, in order to reveal the complete secret key $\mathbf{s}$. All the $n - 1$ partial keys are absolutely necessary for this, as they are distinct and nonoverlapping, i.e., there is no information redundancy among them. From the perspective of the individual agents, the operation is strictly on a need to know basis, which means that, after the completion of the protocol, they gain no additional information that they did not knew already.

\begin{tcolorbox}
	[
		grow to left by = 0.50 cm,
		grow to right by = 0.50 cm,
		colback = WordTurquoiseLighter80!07,	
		enhanced jigsaw,						
		sharp corners,
		toprule = 1.0 pt,
		bottomrule = 1.0 pt,
		leftrule = 0.1 pt,
		rightrule = 0.1 pt,
		sharp corners,
		center title,
		fonttitle = \bfseries
	]
	\centering
	\begin{figure}[H]
		\begin{tikzpicture}[ scale = 0.90 ]
			\begin{yquant}
				nobit AUX_0_0;
				qubits { $IR_0$: \ $\ket{ GHZ_{n} }^{\otimes m}$ } IR_AGENT_0;
				qubit { $OR_0$: \ $\ket{1}$ } OR_AGENT_0;
				nobit AUX_0_1;
				nobit AUX_0_i;
				nobit AUX_i_0;
				qubits { $IR_i$: \ $\ket{ GHZ_{n} }^{\otimes m}$ } IR_AGENT_i;
				qubit { $OR_i$: \ $\ket{1}$ } OR_AGENT_i;
				nobit AUX_i_1;
				nobit AUX_i_n_2;
				nobit AUX_n_2_0;
				qubits { $IR_{n - 2}$: \ $\ket{ GHZ_{n} }^{\otimes m}$ } IR_AGENT_n_2;
				qubit { $OR_{n - 2}$: \ $\ket{1}$ } OR_AGENT_n_2;
				nobit AUX_n_2_1;
				nobit AUX_A_0;
				qubits { $AIR$: \ $\ket{ GHZ_{n} }^{\otimes m}$ } AIR;
				nobit AUX_A_1;
				[ name = Ph0, WordBlueVeryLight, line width = 0.50 mm, label = Initial State ]
				barrier ( - ) ;
				[ draw = WordRed, fill = WordRed ] [ radius = 0.5 cm ] box {\large\sf{H}} OR_AGENT_0;
				[ draw = WordRed, fill = WordRed ] [ radius = 0.5 cm ] box {\large\sf{H}} OR_AGENT_i;
				[ draw = WordRed, fill = WordRed ] [ radius = 0.5 cm ] box {\large\sf{H}} OR_AGENT_n_2;
				[ name = Ph1, WordBlueVeryLight, line width = 0.50 mm, label = Phase 1 ]
				barrier ( - ) ;
				[ draw = GreenLighter1, fill = GreenLighter1 ] [ x radius = 0.7 cm ] box {\large\sf{U}$_{f_{0}}$} (IR_AGENT_0 - OR_AGENT_0);
				[ draw = GreenLighter1, fill = GreenLighter1 ] [ x radius = 0.7 cm ] box {\large\sf{U}$_{f_{i}}$} (IR_AGENT_i - OR_AGENT_i);
				[ draw = GreenLighter1, fill = GreenLighter1 ] [ x radius = 0.7 cm ] box {\large\sf{U}$_{f_{n - 2}}$} (IR_AGENT_n_2 - OR_AGENT_n_2);
				[ name = Ph2, WordBlueVeryLight, line width = 0.50 mm, label = Phase 2 ]
				barrier ( - ) ;
				[ draw = WordRed, fill = WordRed ] [ radius = 0.5 cm ] box {\large\sf{H}$^{\otimes m}$} IR_AGENT_0;
				[ draw = WordRed, fill = WordRed ] [ radius = 0.5 cm ] box {\large\sf{H}$^{\otimes m}$} IR_AGENT_i;
				[ draw = WordRed, fill = WordRed ] [ radius = 0.5 cm ] box {\large\sf{H}$^{\otimes m}$} IR_AGENT_n_2;
				[ draw = WordRed, fill = WordRed ] [ radius = 0.5 cm ] box {\large\sf{H}$^{\otimes m}$} AIR;
				[ name = Ph3, WordBlueVeryLight, line width = 0.50 mm, label = Phase 3 ]
				barrier ( - ) ;
				[ fill = yellow!50 ] [ radius = 0.5 cm ] measure IR_AGENT_0;
				[ fill = yellow!50 ] [ radius = 0.5 cm ] measure IR_AGENT_i;
				[ fill = yellow!50 ] [ radius = 0.5 cm ] measure IR_AGENT_n_2;
				[ fill = yellow!50 ] [ radius = 0.5 cm ] measure AIR;
				[ name = Ph4, WordBlueVeryLight, line width = 0.50 mm, label = Phase 4 ]
				barrier ( - ) ;
				hspace {0.3 cm} IR_AGENT_0;
				output {$\ket{ \mathbf{y}_0 }$} IR_AGENT_0;
				output {$\ket{ \mathbf{y}_i }$} IR_AGENT_i;
				output {$\ket{ \mathbf{y}_{n - 2} }$} IR_AGENT_n_2;
				output {$\ket{ \mathbf{a} }$} AIR;
				\node [ below = 6.25 cm ] at (Ph0) {$\ket{\psi_{0}}$};
				\node [ below = 6.25 cm ] at (Ph1) {$\ket{\psi_{1}}$};
				\node [ below = 6.25 cm ] at (Ph2) {$\ket{\psi_{2}}$};
				\node [ below = 6.25 cm ] at (Ph3) {$\ket{\psi_{3}}$};
				\node [ below = 6.25 cm ] at (Ph4) {$\ket{\psi_{4}}$};
				\node
				[
				maninblack,
				scale = 1.00,
				anchor = center,
				label = { [ label distance = 0.00 cm ] south: Agent$_0$ }
				]
				( ) at ( -2.35, -1.50 ) { };
				\node
				[
				maninblack, shirt = brown!50!gray, skin = brown!70!black,
				scale = 1.00,
				anchor = center,
				label = { [ label distance = 0.00 cm ] south: Agent$_i$ }
				]
				( ) at ( -2.35, -5.10 ) { };
				\node
				[
				maninblack, shirt = green!20!blue, hair = gray!70!black, skin = brown,
				scale = 1.00,
				anchor = center,
				label = { [ label distance = 0.00 cm ] south: Agent$_{n - 2}$ }
				]
				( ) at ( -2.35, -8.90 ) { };
				\node
				[
				alice,
				scale = 1.00,
				anchor = center,
				label = { [ label distance = 0.00 cm ] south: Alice }
				]
				(Alice) at ( -2.35, -11.65 ) { };
				
			\end{yquant}
		\end{tikzpicture}
		\caption{The above figure shows the quantum circuits employed by Alice and her agents. We point out that these circuits are spatially separated, but, due to entanglement, strongly correlated forming a composite system. The state vectors $\ket{\psi_{0}}$, $\ket{\psi_{1}}$, $\ket{\psi_{2}}$, $\ket{\psi_{3}}$ and $\ket{\psi_{4}}$ describe the evolution of the composite system.}
		\label{fig:The Quantum Circuit for the QSA Protocol}
	\end{figure}
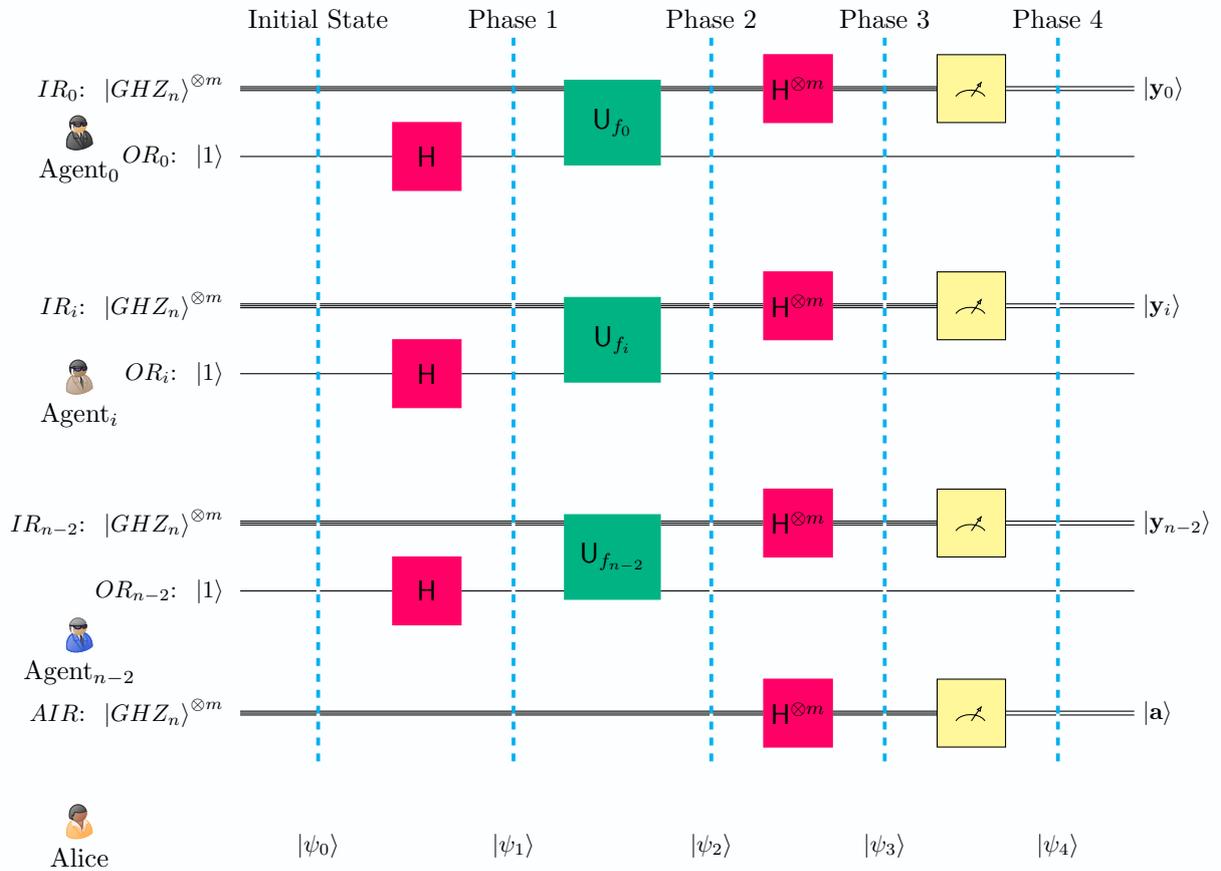
\end{tcolorbox}

The QSA protocol successfully accomplishes this feat, by employing the quantum circuit shown in Figure \ref{fig:The Quantum Circuit for the QSA Protocol}. There, we show the individual quantum circuits employed by Alice and her $n - 1$ agents Agent$_0$, \dots, Agent$_{n - 2}$. Table \ref{tbl:Figure QSA Notations and Abbreviations} explains the abbreviations that are used in the quantum circuit depicted in Figure \ref{fig:The Quantum Circuit for the QSA Protocol}. It is important to emphasize that this is a distributed quantum circuit made up of $n$ individual, spatially separated and private circuits. It is the phenomenon of entanglement that strongly correlates the individual subsircuits, forming in a effect a composite distributed circuit. The state vectors $\ket{\psi_{0}}$, $\ket{\psi_{1}}$, $\ket{\psi_{2}}$, $\ket{\psi_{3}}$ and $\ket{\psi_{4}}$ describe the evolution of the composite system. The $n$ individual subcircuits have obvious similarities, and some important differences, as summarized in Table \ref{tbl:Differences and Similarities of the SubCircuits}. Let us also clarify that for consistency we follow the Qiskit \cite{Qiskit2022} convention in the ordering of qubits, by placing the least significant at the top and the most significant at the bottom.

In our subsequent analytical mathematical description of the QSA game, we use the typical convention of writing the contents of quantum registers in boldface, e.g., $\ket{ \mathbf{x} } = \ket{ x_{ m - 1 } } \dots \ket{ x_{0} }$, for some $m \geq 1$. Moreover, apart from equation (\ref{eq:m Extended General GHZ_n States}), we will make use of the two other well-known formulas given below (see any standard textbook, such as \cite{Nielsen2010} or \cite{Mermin2007}).

\begin{align}
	H \ket{1}
	&=
	\frac{1}{\sqrt{2}}
	\left( \ket{0} - \ket{1} \right)
	=
	\ket{-}
	\label{eq:Ket - Definition}
	\\
	H^{\otimes m} \ket{ \mathbf{x} }
	&=
	\frac{1}{\sqrt{2^m}}
	\sum_{ \mathbf{z} \in \{ 0, 1 \}^m }
	(-1)^{ \mathbf{z \cdot x} } \ket{ \mathbf{z} }
	\ ,
	\label{eq:Hadamard m-Fold Ket x}
\end{align}

where $\ket{ \mathbf{z} } = \ket{ z_{ m - 1 } } \dots \ket{ z_{0} }$ and $\mathbf{z \cdot x}$ is the \emph{inner product modulo} $2$, defined as 

\begin{align} \label{eq:Inner Product Modulo $2$}
	\mathbf{z \cdot x} = z_{ m - 1 } x_{ m - 1 } \oplus \dots \oplus z_{0} x_{0}
	\ .
\end{align}

\begin{tcolorbox}
	[
		grow to left by = 0.00 cm,
		grow to right by = 0.00 cm,
		colback = WordTurquoiseLighter80!07,	
		enhanced jigsaw,						
		sharp corners,
		boxrule = 0.1 pt,
		toprule = 0.1 pt,
		bottomrule = 0.1 pt
	]
	\begin{table}[H]
		\renewcommand{\arraystretch}{1.40}
		\caption{This table contains the notations and abbreviations that are used in Figure \ref{fig:The Quantum Circuit for the QSA Protocol}.}
		\label{tbl:Figure QSA Notations and Abbreviations}
		\centering
		\begin{tabular}
			{
				>{\centering\arraybackslash} m{2.50 cm} !{\vrule width 1.25 pt}
				>{\centering\arraybackslash} m{9.50 cm} 
			}
			\Xhline{7\arrayrulewidth}
			\multicolumn{2}{c}{\textbf{Notations and Abbreviations}}
			\\
			\Xhline{\arrayrulewidth}
			Symbolism
			&
			Explanation
			\\
			$n$
			&
			Number of players (Alice plus her $n - 1$ agents)
			\\
			\Xhline{1\arrayrulewidth}
			\multirow{3}{*}{$m$}
			&
			Length of the secret key $\mathbf{s}$, equal to the
			\\
			&
			number of qubits in the Input Registers
			\\
			&
			of Alice \& every one of her agents
			\\
			\Xhline{1\arrayrulewidth}
			AIR
			&
			Alice's $m$-qubit Input Register
			\\
			\Xhline{1\arrayrulewidth}
			IR$_{ i }$
			&
			The $m$-qubit Input Register of Agent$_i, \ 0 \leq i \leq n - 2$
			\\
			\Xhline{1\arrayrulewidth}
			OR$_{ i }$
			&
			The single-qubit Output Register of Agent$_i, \ 0 \leq i \leq n - 2$
			\\
			\Xhline{7\arrayrulewidth}
		\end{tabular}
		\renewcommand{\arraystretch}{1.0}
	\end{table}
\end{tcolorbox}

\begin{tcolorbox}
	[
		grow to left by = 2.00 cm,
		grow to right by = 0.00 cm,
		colback = WordTurquoiseLighter80!07,	
		enhanced jigsaw,						
		sharp corners,
		boxrule = 0.1 pt,
		toprule = 0.1 pt,
		bottomrule = 0.1 pt
	]
	\begin{table}[H]
		\renewcommand{\arraystretch}{1.40}
		\caption{Differences and similarities among the $n$ subcircuits depicted in Figure \ref{fig:The Quantum Circuit for the QSA Protocol}.}
		\label{tbl:Differences and Similarities of the SubCircuits}
		\centering
		\begin{tabular}
			{
				>{\centering\arraybackslash} m{7.00 cm} !{\vrule width 1.25 pt}
				>{\centering\arraybackslash} m{7.00 cm} 
			}
			\Xhline{7\arrayrulewidth}
			\multicolumn{2}{c}{\textbf{Differences and Similarities}}
			\\
			\Xhline{\arrayrulewidth}
			Differences
			&
			Similarities
			\\
			Alice's circuit lacks Output Register
			&
			All circuits contain an $m$-qubit Input Register
			\\
			\Xhline{1\arrayrulewidth}
			Alice does not apply any function 
			&
			All agents' circuits contain an Output Register
			\\
			\Xhline{1\arrayrulewidth}
			Every agent applies a different function $f_{i}$
			&
			All Output Registers are initialized to $\ket{1}$
			\\
			\Xhline{1\arrayrulewidth}
			\multirow{3}{*}{ }
			&
			All circuits apply the $m$-fold
			\\
			&
			Hadamard transform on their
			\\
			&
			Input Register prior to measurement
			\\
			\Xhline{7\arrayrulewidth}
		\end{tabular}
		\renewcommand{\arraystretch}{1.0}
	\end{table}
\end{tcolorbox}

The initial state $\ket{ \psi_0 }$ of the circuit shown in Figure \ref{fig:The Quantum Circuit for the QSA Protocol} is given by

\begin{align} \label{eq:SQA Phase 0}
	\ket{ \psi_0 }
	=
	\frac{1}{ \sqrt{2^m} }
	\sum_{ \mathbf{x} \in \{ 0, 1 \}^m }
	\ket{\mathbf{x}}_{ A }
	\ket{1}_{ n - 2 }
	\ket{\mathbf{x}}_{ n - 2 }
	\dots
	\ket{1}_{ 0 }
	\ket{\mathbf{x}}_{ 0 }
	\ .
\end{align}

In equation (\ref{eq:SQA Phase 0}), $\ket{\mathbf{x}}_{A}$ designates the contents of Alice's Input Register, $\ket{1}_{i}, \ 0 \leq i \leq n - 2,$ is the state of the agents' Output Registers, and $\ket{\mathbf{x}}_{i}, 0 \leq i \leq n - 2,$ denotes the contents of the Input Registers of the $n - 1$ agents. In what follows, the subscripts $A$ and $0, 1, \dots, n - 2$ are utilized in an effort to distinguish between the local registers of Alice and Agent$_0$, \dots, Agent$_{ n - 2 }$, respectively.

The first phase of the protocol begins when all the agents apply the Hadamard transform to their respective Output Register, driving the system to the next state $\ket{ \psi_1 }$

\begin{align} \label{eq:SQA Phase 1}
	\ket{\psi_1}
	=
	\frac{1}{ \sqrt{2^m} }
	\sum_{ \mathbf{x} \in \{ 0, 1 \}^m }
	\ket{\mathbf{x}}_{ A }
	\ket{-}_{ n - 2 }
	\ket{\mathbf{x}}_{ n - 2 }
	\dots
	\ket{-}_{ 0 }
	\ket{\mathbf{x}}_{ 0 }
	\ .
\end{align}

At this point each of the $n - 1$ agents transmits her secret. Since this is the most important part of the protocol, we explain in detail how this task is implemented. Agent$_{i}, \ 0 \leq i \leq n - 2,$ defines a function that is based on her extended partial secret key $\mathbf{s}_i$, namely

\begin{align} \label{eq:Agent's i Function}
	f_{ i } ( \mathbf{x} )
	=
	\mathbf{s}_i \cdot \mathbf{x} \ , \ 0 \leq i \leq n - 2
	\ .
\end{align}

Agent$_{ i }, \ 0 \leq i \leq n - 2,$ uses function $f_{ i }$ to construct the unitary transform $U_{ f_{ i } }$, which, as is typical of many quantum algorithms, acts on both Output and Input Registers, producing the following output:

\begin{align} \label{eq:Agent's i Oracle - I}
	U_{ f_{ i } } :
	\ket{ y } \ket{ \mathbf{x} }
	\rightarrow
	\ket{ y \oplus f( \mathbf{x} ) } \ket{ \mathbf{x} }
	\ .
\end{align}

Taking into account (\ref{eq:SQA Phase 1}), which asserts that for every agent the state of the Output Register is $\ket{-}$, and (\ref{eq:Agent's i Function}), formula (\ref{eq:Agent's i Oracle - I}) becomes

\begin{align} \label{eq:Agent's i Oracle - II}
	U_{ f_{ i } } :
	\ket{ - } \ket{ \mathbf{x} }
	\rightarrow
	( - 1 )^{ \mathbf{s}_{ i } \cdot \mathbf{x} }
	\ket{ - } \ket{ \mathbf{x} }
	\ .
\end{align}

Hence, the cumulative action of the unitary transforms $U_{f_{i}}, \ 0 \leq i \leq n - 2,$ sends the quantum circuit to the next state:

\begin{align} \label{eq:SQA Phase 2}
	\ket{\psi_2}
	&=
	\frac{ 1 }{ \sqrt{ 2^m } }
	\sum_{ \mathbf{x} \in \{ 0, 1 \}^m }
	\ket{\mathbf{x}}_{ A }
	( - 1 )^{ \mathbf{s}_{ n - 2 } \cdot \mathbf{x} }
	\ket{-}_{ n - 2 }
	\ket{\mathbf{x}}_{ n - 2 }
	\dots
	( - 1 )^{ \mathbf{s}_{ 0 } \cdot \mathbf{x} }
	\ket{-}_{ 0 }
	\ket{\mathbf{x}}_{ 0 }
	\nonumber \\
	&=
	\frac{ 1 }{ \sqrt{ 2^m } }
	\sum_{ \mathbf{x} \in \{ 0, 1 \}^m }
	( - 1 )^{ ( \mathbf{s}_{ n - 2 } \oplus \dots \oplus \mathbf{s}_{ 0 } ) \cdot \mathbf{x} }
	\ket{\mathbf{x}}_{ A }
	\ket{-}_{ n - 2 }
	\ket{\mathbf{x}}_{ n - 2 }
	\dots
	\ket{-}_{ 0 }
	\ket{\mathbf{x}}_{ 0 }
	\nonumber \\
	&\overset{(\ref{eq:Complete Secret as Mod $2$ Sum of Partial Keys})}{=}
	\frac{ 1 }{ \sqrt{ 2^m } }
	\sum_{ \mathbf{x} \in \{ 0, 1 \}^m }
	( - 1 )^{ \mathbf{s} \cdot \mathbf{x} }
	\ket{\mathbf{x}}_{ A }
	\ket{-}_{ n - 2 }
	\ket{\mathbf{x}}_{ n - 2 }
	\dots
	\ket{-}_{ 0 }
	\ket{\mathbf{x}}_{ 0 }
	\ .
\end{align}

At this point, the complete secret key is implicitly encoded in the state of the circuit. It remains to be deciphered by Alice, as explained in the next subsection.

\subsection{Retrieval phase} \label{subsec:Retrieval Phase}

Subsequently, Alice and all her spies apply the $m$-fold Hadamard transformation to their Input Registers. The next state of the circuit is shown below. Please note that henceforth, and in order to make the remaining formulas more readable and understandable, we have chosen to omit the Output Registers; they have served their intended purpose and will no longer be of any use.

\begin{tcolorbox}
	[
		grow to left by = 0.85 cm,
		grow to right by = 0.85 cm,
		colback = white,			
		enhanced jigsaw,			
		sharp corners,
		boxrule = 0.01 pt,
		toprule = 0.01 pt,
		bottomrule = 0.01 pt
	]
	{\small
		\begin{align} \label{eq:SQA Phase 3 - I}
			\ket{\psi_3}
			=
			&\frac{ 1 }{ \sqrt{ 2^m } }
			\sum_{ \mathbf{x} \in \{ 0, 1 \}^m }
			( - 1 )^{ \mathbf{s} \cdot \mathbf{x} }
			H^{ \otimes m } \ket{ \mathbf{x} }_{A}
			H^{ \otimes m } \ket{\mathbf{x}}_{ n - 2 }
			\dots
			H^{ \otimes m } \ket{\mathbf{x}}_{0}
			\nonumber \\
			\overset{(\ref{eq:Hadamard m-Fold Ket x})}{=}
			&\frac{ 1 }{ \sqrt{ 2^m } }
			\sum_{ \mathbf{x} \in \{ 0, 1 \}^m }
			( - 1 )^{ \mathbf{s} \cdot \mathbf{x} }
			\left(
			\frac{ 1 }{ \sqrt{ 2^m } }
			\sum_{ \mathbf{a} \in \{ 0, 1 \}^m }
			( - 1 )^{ \mathbf{a} \cdot \mathbf{x} }
			\ket{ \mathbf{a} }_{A}
			\right)
			\nonumber \\
			&\left(
			\frac{ 1 }{ \sqrt{ 2^m } }
			\sum_{ \mathbf{y}_{ n - 2 } \in \{ 0, 1 \}^m }
			( - 1 )^{ \mathbf{y}_{ n - 2 } \cdot \mathbf{x} }
			\ket{ \mathbf{y}_{ n - 2 } }_{ n - 2 }
			\right)
			\dots
			\left(
			\frac{ 1 }{ \sqrt{ 2^m } }
			\sum_{ \mathbf{y}_{ 0 } \in \{ 0, 1 \}^m }
			( - 1 )^{ \mathbf{y}_{ 0 } \cdot \mathbf{x} }
			\ket{ \mathbf{y}_{ 0 } }_{0}
			\right)
			\nonumber \\
			=
			&\frac{ 1 }{ ( \sqrt{ 2^m } )^{ n + 1 } }
			\sum_{ \mathbf{x} \in \{ 0, 1 \}^m }
			\sum_{ \mathbf{a} \in \{ 0, 1 \}^m }
			\sum_{ \mathbf{y}_{ n - 2 } \in \{ 0, 1 \}^m }
			\dots
			\sum_{ \mathbf{y}_{ 0 } \in \{ 0, 1 \}^m }
			( - 1 )^{ ( \mathbf{s} \oplus \mathbf{a} \oplus \mathbf{y}_{ n - 2 } \oplus \cdots \oplus \mathbf{y}_{ 0 } ) \cdot \mathbf{x} }
			\ket{ \mathbf{a} }_{A}
			\ket{ \mathbf{y}_{ n - 2 } }_{ n - 2 }
			\dots
			\ket{ \mathbf{y}_{ 0 } }_{0}
			\ .
		\end{align}
	}
\end{tcolorbox}

The above formula looks complicated but it can be simplified by invoking an important property of the inner product modulo $2$ operation. If $\ket{ \mathbf{c} } = \ket{ c_{ m - 1 } } \dots \ket{ c_{0} } \neq \ket{ 0 }^{\otimes m}$ is a fixed basis ket, then for \emph{precisely half} of the basis kets $\ket{ \mathbf{x} }$, \ $\mathbf{c} \cdot \mathbf{x}$ will be $0$ and for the \emph{remaining half}, \ $\mathbf{c} \cdot \mathbf{x}$ will be $1$. In the special case where $\ket{ \mathbf{c} } = \ket{ 0 }^{\otimes m}$, then for \emph{every} basis ket $\ket{ \mathbf{x} }$, $\ \mathbf{c} \cdot \mathbf{x} = 0$. Applying this property to equation (\ref{eq:SQA Phase 3 - I}), we conclude that if

\begin{align} \label{eq:Application of the Inner Product Modulo 2 Property - I}
	\mathbf{a} \oplus \mathbf{y}_{ n - 2 } \oplus \cdots \oplus \mathbf{y}_{ 0 } = \mathbf{s} \ ,
\end{align}

then, for each $\mathbf{x} \in \{ 0, 1 \}^m$, the expression $( - 1 )^{ ( \mathbf{s} \oplus \mathbf{a} \oplus \mathbf{y}_{ n - 2 } \oplus \cdots \oplus \mathbf{y}_{ 0 } ) \cdot \mathbf{x} }$ becomes $( - 1 )^{0} = 1$. Therefore, the sum $\sum_{ \mathbf{x} \in \{ 0, 1 \}^m } ( - 1 )^{ ( \mathbf{s} \oplus \mathbf{a} \oplus \mathbf{y}_{ n - 2 } \oplus \cdots \oplus \mathbf{y}_{ 0 } ) \cdot \mathbf{x} }$ equals $2^{m}$. In contrast, when $\mathbf{a} \oplus \mathbf{y}_{ n - 2 } \oplus \cdots \oplus \mathbf{y}_{ 0 } \neq \mathbf{s}$, the sum reduces to $0$. This is typically written in a compact way as

\begin{align} \label{eq:General Inner Product Modulo 2 Property - II}
	\sum_{ \mathbf{x} \in \{ 0, 1 \}^m }^{}
	( - 1 )^{ ( \mathbf{s} \oplus \mathbf{a} \oplus \mathbf{y}_{ n - 2 } \oplus \cdots \oplus \mathbf{y}_{ 0 } ) \cdot \mathbf{x} }
	=
	2^{m} \delta_{ \mathbf{s}, \mathbf{a} \oplus \mathbf{y}_{ n - 2 } \oplus \cdots \oplus \mathbf{y}_{ 0 } }
	\ .
\end{align}

In view of (\ref{eq:General Inner Product Modulo 2 Property - II}), we may express state $\ket{\psi_3}$ more succinctly as

\begin{align} \label{eq:SQA Phase 3 - II}
	\ket{\psi_3}
	&=
	\frac{ 1 }{ ( \sqrt{ 2^m } )^{ n - 1 } }
	\sum_{ \mathbf{a} \in \{ 0, 1 \}^m }
	\sum_{ \mathbf{y}_{ n - 2 } \in \{ 0, 1 \}^m }
	\dots
	\sum_{ \mathbf{y}_{ 0 } \in \{ 0, 1 \}^m }
	\ket{ \mathbf{a} }_{A}
	\ket{ \mathbf{y}_{ n - 2 } }_{ n - 2 }
	\dots
	\ket{ \mathbf{y}_{ 0 } }_{0}
	\ .
\end{align}

The fundamental property of the QSA protocol, as encoded in equations (\ref{eq:General Inner Product Modulo 2 Property - II}) and (\ref{eq:SQA Phase 3 - II}) states that the contents of the Input Registers of Alice and all her $n - 1$ agents \emph{can not vary completely freely and independently}. The presence of tuples entangled in the $\ket{ GHZ_{n} }$ state during the initialization of the quantum circuit, has manifested itself in state $\ket{\psi_3}$ in what we call the \textbf{Fundamental Correlation Property}. This property asserts that in \emph{each term} of the linear combination described by $\ket{\psi_3}$, the states $\ket{ \mathbf{a} }_{A}, \ket{ \mathbf{y}_{ n - 2 } }_{ n - 2 }, \dots, \ket{ \mathbf{y}_{ 0 } }_{0}$ of the $n$ players' Input Registers are correlated by the following constraint:

\begin{align} \label{eq:QSA Fundamental Correlation Property}
	\mathbf{a}
	\oplus
	\mathbf{y}_{ n - 2 }
	\oplus \cdots \oplus
	\mathbf{y}_{ 0 } = \mathbf{s}
	\ .
\end{align}

The quantum part of the QSA protocol is completed when all players, i.e., Alice and her secret agents Agent$_0$, \dots, Agent$_{ n - 2 }$ measure their Input Registers, which results in the final state $\ket{\psi_4}$ of the quantum circuit.

\begin{align}
	\label{eq:QSA Final Measurement}
	\ket{\psi_4}
	=
	\ket{ \mathbf{a} }_{A}
	\ket{ \mathbf{y}_{ n - 2 } }_{ n - 2 }
	\dots
	\ket{ \mathbf{y}_{ 0 } }_{0}
	\ , \quad \text{for some} \quad
	\mathbf{a}, \mathbf{y}_{ 0 }, \dots, \mathbf{y}_{ n - 2 } \in \{ 0, 1 \}^m
	\ ,
\end{align}

where $\mathbf{a}, \mathbf{y}_{ 0 }, \dots, \mathbf{y}_{ n - 2 }$ are correlated via (\ref{eq:QSA Fundamental Correlation Property}). The unique advantage of entanglement has led to this situation: although the contents of each of the $n$ Input Registers may deceptively seem completely random to each player, in fact they are not. The distributed quantum circuit of Figure \ref{fig:The Quantum Circuit for the QSA Protocol}, considered as a composite system, ensures that the final contents of the Input Registers satisfy the Fundamental Correlation Property, as expressed by (\ref{eq:QSA Fundamental Correlation Property}).

\begin{tcolorbox}
	[
		grow to left by = 0.00 cm,
		grow to right by = 0.00 cm,
		colback = yellow!03!white,			
		enhanced jigsaw,					
		sharp corners,
		toprule = 1.0 pt,
		bottomrule = 1.0 pt,
		leftrule = 0.1 pt,
		rightrule = 0.1 pt,
		sharp corners,
		center title,
		fonttitle = \bfseries
	]
	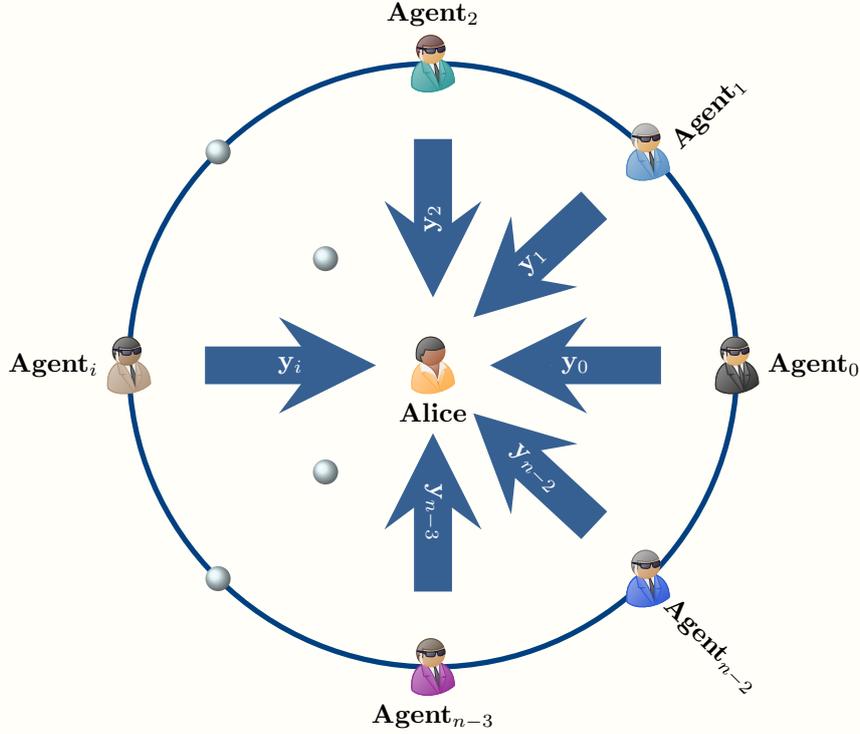
\begin{figure}[H]
		\centering
		\begin{tikzpicture} [scale = 1.00]
			\def \n {8}
			\def \Angle {360 / \n}
			\def \Radius {4}
			\node
			[
			shade, top color = WordBlueDarker25, bottom color = black, rectangle, text width = 9.00 cm, align = center
			] ( Label ) at ( 0.0, 7.00 )
			{ \color{white} \textbf{THE CLASSICAL CHANNEL}\\
				The $n - 1$ spatially distributed agents send to Alice through the classical channel the measurements $\mathbf{y}_{ 0 }, \dots, \mathbf{y}_{ n - 2 }$ of their Input Registers.};
			\draw [ line width = 2.00 pt, MyBlue ] ( 0, 0 ) circle [ radius = \Radius cm ];
			\node
			[
			maninblack,
			scale = 1.50,
			anchor = center,
			label = { [ label distance = 0.00 cm ] east: \textbf{Agent$_0$} }
			]
			( ) at ( { \Radius * cos(0 * \Angle) }, { \Radius * sin(0 * \Angle) } ) { };
			\node
			[
			maninblack, shirt = WordBlueDark, hair = gray,
			scale = 1.50,
			anchor = center,
			label = { [ rotate = \Angle, label distance = 0.00 cm ] east: \textbf{Agent$_1$} }
			]
			( ) at ( { \Radius * cos(1 * \Angle) }, { \Radius * sin(1 * \Angle) } ) { };
			\node
			[
			maninblack, shirt = GreenTeal, hair = red!20!black,
			scale = 1.50,
			anchor = center,
			label = { [ label distance = 0.00 cm ] north: \textbf{Agent$_2$} }
			]
			( ) at ( { \Radius * cos(2 * \Angle) }, { \Radius * sin(2 * \Angle) } ) { };
			\node [ shade, shading = ball, ball color = WordAquaLighter80, circle ] () at ( { \Radius * cos(3 * \Angle) }, { \Radius * sin(3 * \Angle) } ) {};
			\node
			[
			maninblack, shirt = brown!50!gray, skin = brown!70!black,
			scale = 1.50,
			anchor = center,
			label = { [ label distance = 0.00 cm ] west: \textbf{Agent$_i$} }
			]
			( ) at ( { \Radius * cos(4 * \Angle) }, { \Radius * sin(4 * \Angle) } ) { };
			\node [ shade, shading = ball, ball color = WordAquaLighter80, circle ] () at ( { \Radius * cos(5 * \Angle) }, { \Radius * sin(5 * \Angle) } ) {};
			\node
			[
			maninblack, shirt = red!50!blue, hair = brown!30!black, skin = brown,
			scale = 1.50,
			anchor = center,
			label = { [ label distance = 0.00 cm ] south: \textbf{Agent$_{n - 3}$} }
			]
			( ) at ( { \Radius * cos(6 * \Angle) }, { \Radius * sin(6 * \Angle) } ) { };
			\node
			[
			maninblack, shirt = green!20!blue, hair = gray!70!black, skin = brown,
			scale = 1.50,
			anchor = center,
			label = { [ rotate = - \Angle, label distance = 0.00 cm ] right: \textbf{Agent$_{n - 2}$} }
			]
			( ) at ( { \Radius * cos(7 * \Angle) }, { \Radius * sin(7 * \Angle) } ) { };
			\node
			[
			alice,
			scale = 1.50,
			anchor = center,
			label = { [ label distance = 0.00 cm ] south: \textbf{Alice} }
			]
			(Alice) at (0.0, 0.0) { };
			\begin{scope}[on background layer]
				\foreach \angle / \index in { 0/0, 45/1, 90/2, 270/n-3, 315/n-2 }
				\draw [ WordBlueDarker25, <-, >=stealth, line width = 5.0 mm ] 
				( { 0.75 * cos(\angle) }, { 0.90 * sin(\angle) } ) --
				( { 3.00 * cos(\angle) }, { 3.00 * sin(\angle) } )
				node [ midway, text = white, rotate = \angle ] {$\mathbf{y}_{\index}$};
				\foreach \angle in { 135, 225 }
				\node [ shade, shading = ball, ball color = WordAquaLighter80, circle ] () at ( { \Radius * 0.5 * cos(\angle) }, { \Radius * 0.5 * sin(\angle) } ) {};
				\draw [ WordBlueDarker25, <-, >=stealth, line width = 5.0 mm ]
				( { 0.75 * cos(180) }, { 0.75 * sin(180) } ) --
				( { 3.00 * cos(180) }, { 3.00 * sin(180) } )
				node [ midway, text = white ] {$\mathbf{y}_{i}$} ;
			\end{scope}
		\end{tikzpicture}
		\caption{The above figure visualizes the conclusion of the QSA protocol when the $n - 1$ spatially distributed agents in the spy network send to Alice through the classical channel the final measurements $\mathbf{y}_{ 0 }, \dots, \mathbf{y}_{ n - 2 }$ of their Input Registers.} \label{fig:The Agents Transmits to Alice their Measurements}
	\end{figure}
\end{tcolorbox}

One final step remains. Agent$_0$, \dots, Agent$_{ n - 2 }$ must all send the contents of their Input registers $\mathbf{y}_{ 0 }, \dots, \mathbf{y}_{ n - 2 }$, respectively, to Alice, so as to allow Alice to uncover the big secret $\mathbf{s}$. This can be achieved by communicating through the classical channel. Figure \ref{fig:The Agents Transmits to Alice their Measurements} gives a mnemonic visualization of the conclusion of the QSA protocol.

The use of a public channel by the agents to broadcast their measurements will not compromise the security of the protocol for two reasons. First, the transmitted information $\mathbf{y}_{ i }, \ 0 \leq i \leq n - 2,$ is completely unrelated to the extended partial secret $\mathbf{s}_{ i }$. The latter cannot be recovered from the former. Secondly, in the general case, even if Eve combines all the measurements $\mathbf{y}_{ 0 }, \dots, \mathbf{y}_{ n - 2 },$ she still needs $\mathbf{a}$ in order to discover the secret message $\mathbf{s}$. There is of course the special case where $\mathbf{a} = \mathbf{0}$. In such a case Even has all the information she needs to find the secret message $\mathbf{s}$, although she might not know it, i.e., she might have no way to know that Alice's measurement is $\mathbf{0}$. Thus, to secure our protocol from this eventuality, we dictate that Alice should request the repetition of the whole process in the event that the contents of her Input Registers are all zero after the final measurement.

\begin{tcolorbox}
	[
		grow to left by = 0.00 cm,
		grow to right by = 0.00 cm,
		colback = white,			
		enhanced jigsaw,			
		sharp corners,
		boxrule = 0.01 pt,
		toprule = 0.01 pt,
		bottomrule = 0.01 pt
	]
	\begin{figure}[H]
		\centering
		\includegraphics[ scale = 0.45, angle = 90, trim = {0 0 0cm 0}, clip ]{"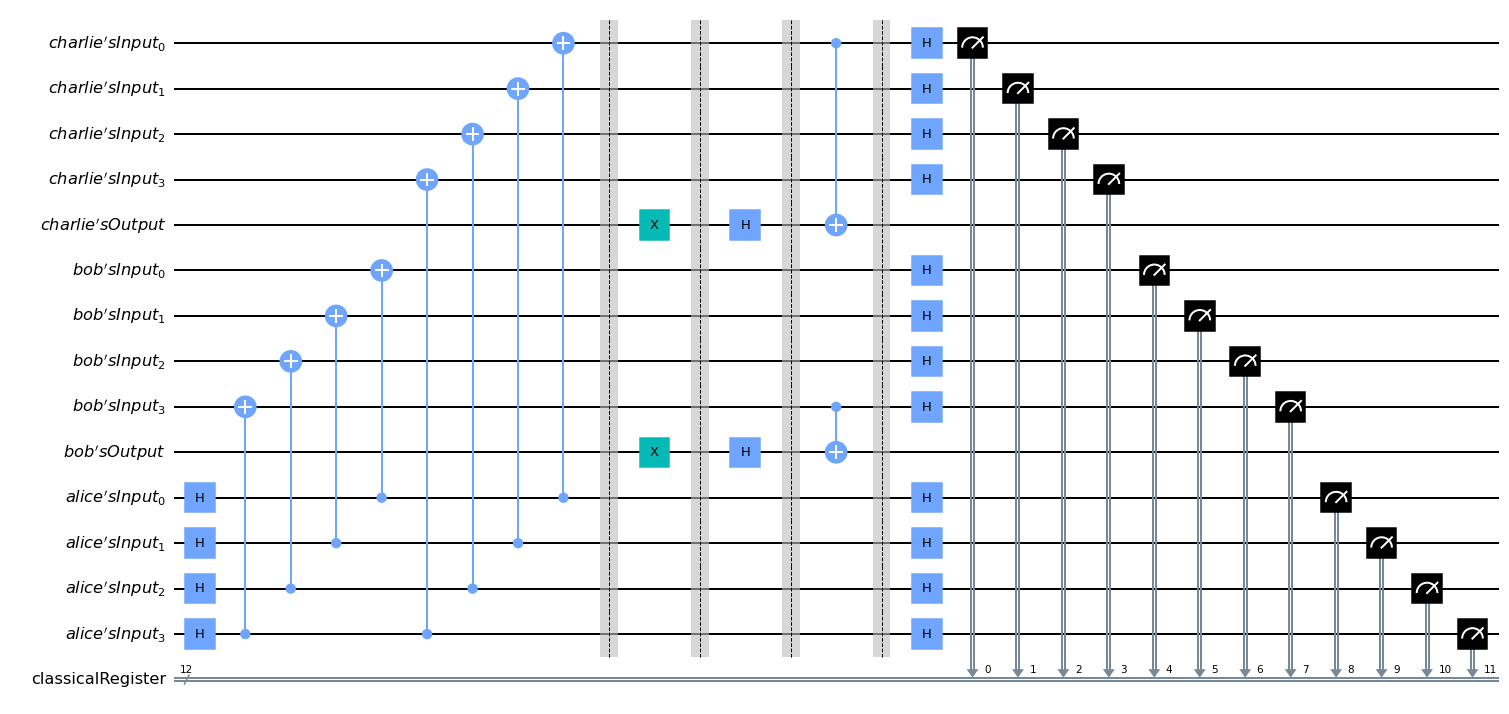"}
		\caption{A toy scale quantum circuit simulating the QSA protocol, as applied to the spymaster Alice and her two agents Bob and Charlie.}
		\label{fig:QSA Example Quantum Circuit}
	\end{figure}
\end{tcolorbox}

\begin{tcolorbox}
	[
		grow to left by = 3.50 cm,
		grow to right by = 0.00 cm,
		colback = white,			
		enhanced jigsaw,			
		sharp corners,
		boxrule = 0.01 pt,
		toprule = 0.01 pt,
		bottomrule = 0.01 pt
	]
	\begin{figure}[H]
		\centering
		\includegraphics[ scale = 0.36, angle = 90, trim = {222.4cm 0cm 211.2cm 8cm}, clip ]{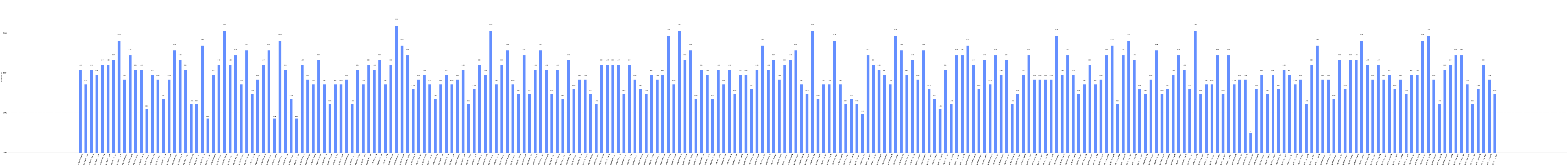}
		\caption{Some of the possible measurements and their corresponding probabilities for the circuit of Figure \ref{fig:QSA Example Quantum Circuit}.}
		\label{fig:QSA Example Measurement Outcomes}
	\end{figure}
\end{tcolorbox}

\section{A toy scale example demonstrating the QSA protocol} \label{sec:QSA Protocol Example}

In this section we present a toy scale example that should be viewed as a proof of concept about the viability of the QSA protocol. The resulting quantum circuit is illustrated in Figure \ref{fig:QSA Example Quantum Circuit}. It was designed and simulated using IBM's \emph{Qiskit} open source SDK (\cite{Qiskit2022}) and, in particular, the Aer provider utilizing the high performance \emph{qasm} simulator for simulating quantum circuits \cite{Qasm2022}. The measurements, of which only a small portion is shown in Figure \ref{fig:QSA Example Measurement Outcomes}, as their sheer number makes their complete visualization inexpedient, along with their corresponding probabilities were obtained by running the qasm simulator for 4096 shots.

In the current example, Alice's network consists of just two agents, nonother than Bob and Charlie. All of them are in different locations. Bob's partial secret key is $\mathbf{p}_{ B } = 10$ and Charlie's partial secret key is $\mathbf{p}_{ C } = 01$. Hence, their extended partial secret keys are $\mathbf{s}_{ B } = 1000$ and $\mathbf{s}_{ C } = 0001$, and the complete secret key that ALice must uncover is $\mathbf{s} = 1001$. As we clarified above, the local quantum circuit of Figure \ref{fig:QSA Example Quantum Circuit} is best considered to be a proof of concept. This is because, at present, we are unable simulate in Qiskit the fact that Alice, Bob, and Charlie are spatially separated. An actual implementation of the QSA protocol would result in a distributed quantum circuit and not a local one as shown in Figure \ref{fig:QSA Example Quantum Circuit}. Furthermore, we are also unable to directly specify a trusted third party source that generates the entangled GHZ triples, although Qiskit provides the ability to initialize the quantum circuit in specific initial state. In any case, we have opted for circuit itself to create the GHZ triples. Hence, these assumptions cannot be accurately reflected in the quantum circuit of Figure \ref{fig:QSA Example Quantum Circuit} and this example should be considered a faithful representation of a real life scenario.

With all the above observations duly noted, we may verify that this simulation is indeed a localized version of the blueprint for the QSA protocol, as shown in Figure \ref{fig:The Quantum Circuit for the QSA Protocol}. The final measurements by Alice, Bob and Charlie will produce one of the $2^{8} = 256$ equiprobable outcomes. Showing all these outcomes would result in an unintelligible figure, so we have opted for depicting only some of them in Figure \ref{fig:QSA Example Measurement Outcomes}. This figure also shows the corresponding probabilities for each outcome; it should not come as a surprise that they are not shown to be equiprobable, as theory expects, since the figure has resulted from a simulation run for 4096 shots. The important thing though is that every possible outcome satisfies the Fundamental Correlation Property and verifies equation (\ref{eq:QSA Fundamental Correlation Property}). Therefore, ignoring the unlikely case that Alice measures $\mathbf{a} = 0000$ in her Input Register, Bob and Charlie, after measuring their Input Registers and obtaining $\mathbf{y}_{ B }$ and $\mathbf{y}_{ C }$, respectively, they only have to send their measurements to Alice so that she can uncover the secret key.

\section{Security analysis of the QSA protocol} \label{sec:Security Analysis of the QSA Protocol}

\subsection{Assumptions} \label{subsec:Assumptions}

In the this section, we shall focus on analyzing several different attack strategies, that a malicious individual, namely Eve, can incorporate against our protocol, with the goal of acquiring a piece of the secret message or in the worst-case scenario the complete message. This will allow us to establish the security of our protocol and its viability in practical applications. However, before we start with our analysis, it is crucial to first clarify two fundamental assumptions that we take for granted and serve as the basis of our security claims. 

We begin by stating the first and most basic assumption, namely that quantum theory is correct and that we can use quantum mechanics to make accurate predictions about measurement outcomes. The reasoning behind this assumption is quite obvious, due to the fact that if the underlying theory was false in one way or another, certain features of quantum mechanics, such as the no-cloning theorem \cite{wootters1982single}, the monogamy of entanglement \cite{coffman2000distributed} or nonlocality \cite{brunner2014bell}, which are vital for any quantum cryptographic protocol, would not apply and thus, it would have been impossible to create a secure protocol.

The second assumption that we adopt is that quantum theory is complete and there are no other special properties or phenomena of quantum mechanics that we don't know. This means that Eve's movements are restricted by the laws of physics and she can not go beyond what is possible with quantum mechanics, in order to acquire more information from her targets. This assumption by its very nature is not perfect, as the question regarding the completeness of quantum mechanics is still unresolved. But the combination of the correctness of quantum mechanics, along with the requirement that free randomness exists, implies that any future extension of quantum theory, will not improve the predictive abilities of any player \cite{colbeck2011no}.

\subsection{Intercept and resend attack} \label{subsec:Intercept & Resend Attack}

We start our security analysis by inspecting the first attack strategy, which of course is the most basic and intuitive type of an individual attack, known as intercept and resend or (I\&R) attack. The main idea of this strategy is for Eve to get hold of each photon coming from Alice or whoever is responsible for the distribution of the GHZ tuples to the rest of the players at the beginning of the protocol. Afterwards, Eve proceeds to measure them on some predefined basis and based on the result, to prepare a new photon and send it to the intended recipient. For this attack, it is rather obvious that in any of the aforementioned possible scenarios our protocol can be used, the GHZ tuples during the distribution phase of the protocol, do not carry any information as regards the nature of the secret message. Thus, our SQA protocol is secure against this attack strategy.

\subsection{PNS attack} \label{subsec:PNS Attack}

The next attack strategy, known as the Photon Number Splitting attack or (PNS) for short, was first introduced by Huttner et al. \cite{huttner1995quantum} and further discussed and analyzed by L{\"u}tkenhaus and Brassard et al. in \cite{lutkenhaus2000security, brassard2000limitations}. Today, it is considered as one of the most effective attack strategies that Eve can use against any protocol This is because it exploits the fact that our current detectors are not $100\%$ efficient and our photon sources do not emit single-photon signals all the time, meaning that there is a possibility for a photon source to produce multiple identical photons instead of only one. Therefore, in a realistic scenario, Eve can intercept these pulses coming from the player or the source responsible for the distribution of the GHZ tuples, take one photon from the multi-photon pulse and send the remaining photon(s) to their legitimate recipient undisturbed. In this scenario, Eve once again will not be able to acquire any information regarding the secret message or the random binary strings that will be used to unlock the secret key. This can be explained from the inherent nature of the QSA protocol, which leads to the creation of seemingly random binary strings during the final phase, when all players apply the final $m$-fold Hadamard transform to their corresponding Input Registers. This means that, if we assume that a tuple in the $\ket{GHZ_{ n + 1 }}$ state is created instead of a tuple in the $\ket{GHZ_{ n }}$ state, this $n + 1$-tuple will correspond to the $n$ players plus Eve. Accordingly, during the measurement phase the results would be

\begin{align}
	\label{eq:QSA$_n+1$ Final Measurement PNS attack}
	\ket{\psi_4}
	=
	\ket{ \mathbf{a} }_{A}
	\ket{ \mathbf{y}_{ n - 1 } }_{ E }
	\ket{ \mathbf{y}_{ n - 2 } }_{ n - 2 }
	\dots
	\ket{ \mathbf{y}_{ 0 } }_{0}
	\ , \quad \text{for some} \quad
	\mathbf{a}, \mathbf{y}_{ 0 }, \dots, \mathbf{y}_{ n - 1 } \in \{ 0, 1 \}^m
	\ ,
\end{align}

instead of the anticipated

\begin{align}
	\label{eq:QSA$_n$ Final Measurement PNS attack}
	\ket{\psi_4}
	=
	\ket{ \mathbf{a} }_{A}
	\ket{ \mathbf{y}_{ n - 2 } }_{ n - 2 }
	\dots
	\ket{ \mathbf{y}_{ 0 } }_{0}
	\ , \quad \text{for some} \quad
	\mathbf{a}, \mathbf{y}_{ 0 }, \dots, \mathbf{y}_{ n - 2 } \in \{ 0, 1 \}^m
	\ .
\end{align}

In such a situation, Eve can be considered as an extra player and, thus, her ability to acquire any extra information about the other players' measurement is, like all the other players, nonexistent.

\subsection{Blinding attack} \label{subsec:Blinding Attack}

Finally, we conclude our security analysis with the blinding attack. During this attack strategy, Eve, instead of trying to intercept the GHZ tuples, she blocks and destroys them entirely before they reach the intended players. Then she proceeds to create her own set of GHZ tuples, with a proper ancilla state in each tuple, and then distributes them to the players. From this description, it is obvious that in order for this particular type of attack to work, the entity responsible for the creation and distribution of the GHZ tuples must be a third party source and not a player. Therefore, during this attack Eve will have a full set of tuples in the $\ket{GHZ_{ n + 1 }}$ state, instead of the aforementioned smaller number of tuples in the $\ket{GHZ_{ n + 1 }}$ state acquired exploiting the inefficiency of our current photon sources, during the PNS attack. However, once again the scenario is similar to the PNS attack, meaning that Eve will be considered as an extra player, and in that case she will again be unable to acquire any information regarding the secret message.

\section{Discussion and conclusions} \label{sec:Discussion and Conclusions}

In this article, we have introduced a new problem in the literature of cryptographic protocols, which we call the Quantum Secret Aggregation problem. We have given a solution to the aforementioned problem that is based on the use of maximally entangled GHZ tuples. These are uniformly distributed among the players, which include the spymaster Alice and her network of agents, all of them being in different locations. We conducted a detailed analysis of the proposed protocol and, subsequently, illustrated its use with a toy scale example involving Alice and her two agents Bob and Charlie. Our presentation has been completely general in the sense that number of players can increase as needed, and the players are assumed to be spatially separated. It is clear that the same protocol can immediately accommodate groups of players that are in the same region of space.

In closing, we point out that the security of our protocol is attributed to its entanglement based nature. For instance, Entanglement Monogamy precludes the entanglement of a maximally entangled tuple with any other qubit. This nullifies Eve;s attempts at gaining information by trying to entangle a qubit of the GHZ tuples used in our protocol, during the transmission of the GHZ tuples to the players.

\bibliographystyle{ieeetr}
\bibliography{QSA}

\end{document}